  \documentclass[12pt]{article}
  \usepackage{amsfonts}
  
\catcode`@=11
\def\secteqno{\@addtoreset{equation}{section}%
\def\theequation{\thesection.\arabic{equation}}}
\catcode`@=12
\secteqno
\newcommand{\be}{\begin{equation}}
\newcommand{\ee}{\end{equation}}
\newcommand{\bea}{\begin{eqnarray}}
\newcommand{\eea}{\end{eqnarray}}
\newcommand{\n}{\nonumber \\}
\newcommand{\bref}[1]{(\ref{#1})}
\newcommand{\nn}{\nonumber}

\begin{document}
\thispagestyle{empty}
\vfill
\hfill June 25, 2003\par
\hfill KEK-TH-896 \par
\vskip 20mm
\begin{center}
{\Large\bf Noncommutative Superspace, Supermatrix \\
and Lowest Landau Level}\par
\vskip 6mm
\medskip

\vskip 10mm
{\large Machiko\ Hatsuda$^{1)}$ $^{2)}$, Satoshi Iso$^{1)}$ 
 and Hiroshi Umetsu$^{1)}$ }\par
\medskip
$^{1)}$
{\it 
Theory Division,\ High Energy Accelerator Research Organization (KEK),\\
Tsukuba,\ Ibaraki,\ 305-0801, Japan }\\
$^{2)}$
{\it
Urawa University, Saitama, Saitama, 336-0974, Japan}\\
\medskip\medskip
{\small\sf E-mails:\ mhatsuda@post.kek.jp, satoshi.iso@kek.jp, 
umetsu@post.kek.jp} 
\medskip
\end{center}
\vskip 10mm
\begin{abstract}
By using graded (super) Lie algebras, we can construct noncommutative
superspace on curved homogeneous manifolds. In this paper, we  take 
a flat limit to obtain flat noncommutative superspace. 
We particularly consider $d=2$ and $d=4$ superspaces based on the 
graded Lie algebras $osp(1|2)$, $su(2|1)$ and $psu(2|2)$. 
Jacobi identities of supersymmetry algebras 
and associativities of star products are automatically satisfied.
Covariant derivatives which commute with supersymmetry generators 
are obtained and chiral constraints can be imposed. 
We also discuss that
these noncommutative superspaces can be understood as 
constrained systems analogous to the lowest Landau level system.
\end{abstract} 
\par\par
\newpage
\setcounter{page}{1}
\parskip=7pt
\section{ Introduction}\par
Deformation of superspace by introducing noncommutativity 
has attracted interests recently.
In papers \cite{OV,Seiberg,Peter}, it is discussed that
the background of the RR field strengths 
(in the first two papers, graviphotons) 
in string theory 
gives rise to non anti-commutative fermionic coordinates.
(In this paper, we use the word noncommutative
for the non anti-commutativity of the fermionic coordinates
unless there is any confusion.)
This phenomenon is similar to the well-known case of the
string theory in the NS-NS two form $B$ background, where 
the bosonic space-time coordinates become noncommutative
\cite{shomerus,SW}.
\par
Noncommutative superspace has been  already studied by several 
papers from the field theoretic approach.
In the paper \cite{schwarz}, they discussed a possibility that 
the anticommutators of fermionic variables are written  
in terms of the space-time
coordinates so that the space-time can be generated as a composite 
of fermions.
Quantum deformations of the Poincare supergroup were
also considered in \cite{Kosinski} and then it was 
shown that the chiral
operators are not closed under star products with fermionic
noncommutativity \cite{Ferrara}.
In the paper\cite{Klemm}, general forms of the deformed superspace
in $d=4$
are discussed by imposing the covariance under supertranslations,
Jacobi identity and closure of chiral superfields under the
star product. 
${\cal N}=1/2$ supersymmetry in $d=4$ was proposed very recently
\cite{Seiberg} and its radiative corrections are also 
studied \cite{Rey}.

\par
There is another approach to study the noncommutative superspace
from supermatrix models based on graded Lie algebras.
This approach is a natural extension of constructing bosonic
noncommutative space based on matrix models. 
In addition to an advantage that the system can be 
realized in terms of finite matrices, 
this matrix model approach is especially useful to study gauge theories
on noncommutative space because noncommutative space-time
and gauge fields on it are unified by single matrices.
In this approach, constructions of the 
open Wilson lines or background independence
of the noncommutative gauge theories become manifest
\cite{AIIKKT, IIKK, Seiberg2}.
Generalizations to supermatrices 
were first investigated in the paper
by H.Grosse, C.Klimcik and P.Presnajder\cite{Grosse}.
These authors studies 
supermatrix models based on the $osp(1|2)$ graded Lie algebra
and constructed supersymmetric actions for scalar multiplets
on two-dimensional fuzzy supersphere.
They further studied noncommutative de Rham complex and forms
in another paper \cite{Grosse2}.
Gauge theories on noncommutative superspace can be also 
constructed \cite{HIU22}.
Recently the concept of noncommutative superspace based on a
supermatrix  was also introduced 
in proving the Dijkgraaf-Vafa conjecture as 
the large N reduction by H.Kawai, T.Kuroki and T.Morita 
\cite{Kawai}, which motivated us to start the present work.

In this paper, we first construct noncommutative superspace based on
graded (super) Lie algebras and then take their flat limits to
obtain flat noncommutative superspace. 
Though this approach is 
restricted to construct noncommutative superspace with Lie algebraic
structures, there is 
an advantage that the symmetry is manifest and that Jacobi identities of 
the algebras and  associativity of star products are
automatically satisfied. 
We also construct covariant
derivatives and impose chiral constraints.
\par
Noncommutative superspace can be also understood as a constrained
system analogous to the lowest Landau level system of particles 
moving in a constant magnetic field. 
For superparticles moving in a magnetic field on superspace,
we can introduce two mutually
commutative sets of operators, covariant derivatives and 
guiding center coordinates. By imposing the lowest Landau level
constraint, superspace coordinates become noncommutative. 
Such noncommutativity on superspace was first discussed in
\cite{schwarz}. Our approach is  similar to theirs in two points;
one is to consider de Sitter algebra instead of the
Poincare algebra and the other is to derive  noncommutativity
through Dirac brackets.
\par
We will here explain a basic idea to induce fermionic 
noncommutativity  by considering 
a particle in  a generalized magnetic field on superspace. 
We first introduce supercovariant derivatives; 
\begin{eqnarray}
&& {\bf D}_{i} = {\bf D}_i^0 - e A_i(x,\theta) \label{1.1}\\
&& D_{\alpha} =  D_{\alpha}^0 - e A_{\alpha}(x, \theta) \label{1.2}.
\end{eqnarray}
${\bf D}_i$ are bosonic covariant derivatives in space-time directions
and $D_{\alpha}$ are fermionic ones  in grassmannian directions. 
In the following we write them together as 
$D_I =({\bf D}_i, D_{\alpha})$. 
These derivatives are familiar in supersymmetric gauge theories
\cite{sohnius} or in supergravities \cite{1001}.
They transform covariantly under local gauge transformations
on superspace. 
If the derivatives without gauge fields $D_I^0$ satisfy
\begin{equation}
[D_I^0, D_J^0 \} = i {T_{IJ}}^K D_K^0,
\end{equation}
we can define field strengths as
\begin{equation}
i F_{IJ} =  [D_I, D_J \} - i {T_{IJ}}^K D_K.
\label{superstrength}
\end{equation}
Now we consider a superspace with a constant magnetic field, 
namely, with a constant $F_{IJ}$ background
\footnote{The field strengths must satisfy the (modified) Bianchi
identity and the background $F_{IJ}$ are generally dependent on
superspace coordinates because of the nonvanishing ${T_{IJ}}^K$. }. 
This is analogous
to the two-dimensional system in a constant magnetic field.
A particle moving in a strong magnetic field is restricted in 
the lowest Landau level and the guiding center coordinates become
noncommutative. In other words, the lowest Landau level
condition can be made second class by imposing 
$D_x =D_y=0$ and the Dirac brackets between 
coordinates become noncommutative. 
In a magnetic field on superspace, by generalizing  the
bosonic second class constraints to include fermionic parts of the
covariant derivatives, we can evaluate Dirac brackets for 
the superspace coordinates. In general this leads to the 
noncommutative algebras for the superspace coordinates.
In this paper, we actually show it explicitly for some examples.
\par
This way to generate fermionic noncommutativity becomes more  
important if we consider gauge theories on superspace
based on supermatrix models. 
In ordinary matrix models of the IKKT type \cite{IKKT}, 
we expand the basic 
bosonic degrees of freedom $A_i$ around some classical
solution $A_i^{cl}$; $A_i=A_i^{cl} + a_i$.
The classical background $A_i^{cl}$ 
defines the background space-time and $a_i$ are 
interpreted as gauge fields on this space-time.
If the classical solutions  are noncommutative, 
we can obtain noncommutative gauge theories \cite{AIIKKT, IIKK}.
This unification of space-time and the gauge field into a single 
matrix $A_i$
can be generalized to a unification of noncommutative 
superspace coordinate $\theta_{\alpha}$
and fermionic gauge field $a_{\alpha}$
into a single fermionic matrix 
$A_{\alpha}= \theta_{\alpha} + a_{\alpha}.$
In order to generate such a fermionic  background, we need to 
consider supermatrix models~\cite{supermatrix, HIU22} 
whose components are again supermatrices. 
The fermionic noncommutativity can be generated 
dynamically as the background of the unified fermionic matrix $A_{\alpha}$.
We want to discuss this issue in a forthcoming paper.

\par
In section 2, as a warming up exercise, we briefly review
the noncommutative (fuzzy) sphere based on  $su(2)$ algebra 
and its flat limit by the In$\ddot{\rm o}$n$\ddot{\rm u}$-Wigner
contractions. 
The noncommutativity on the fuzzy sphere can
be understood as the noncommutativity of the coordinates of a particle 
restricted in the lowest Landau level 
on the commutative sphere in a monopole background.
After taking a flat limit, following a general formalism of 
left and right  $SU(2)$  multiplications on group manifold,
we can construct mutually commutative set of covariant derivatives 
and guiding center coordinates. Though it is a well-known fact,
we explain it in details to clarify the origin of noncommutativity
as a constrained system.
\par
In section 3, we first construct a noncommutative curved superspace
based on a graded Lie algebra $osp(1|2)$ and take its 
flat limit. To make a flat limit together with a fermionic
noncommutativity, we need to take an asymmetric scaling
limit of two grassmannian coordinates. In this limit, only half of 
the supersymmetry generators can generate space-time
supersymmetry. The other half generates non-dynamical supersymmetry.
We can impose a chiral constraint to remove the latter half.
After imposing the chiral constraint, 
the system becomes
essentially a one-dimensional system.
In the latter half of the section, we consider the noncommutative
superspace as a constrained system. 
We first 
obtain mutually commutative sets of operators
for superparticle in a constant magnetic field on superspace; 
super covariant derivatives
and super guiding center coordinates. Then by imposing the lowest Landau
level conditions we obtain the noncommutative relations for the
superspace coordinates.
\par 
In section 4, we consider a noncommutative superspace based on
$su(2|1)$ algebra. This gives again two-dimensional supersphere
but with twice as many as supercharges, i.e. four.
We again need to
take an asymmetric scaling for grassmannian coordinates
when we take a flat limit.
All procedures can be performed in parallel to the $osp(1|2)$ case.
\par
In section 5, we consider four dimensional superspace based on
$psu(2|2)$ superalgebra. This gives a coset supermanifold
$PSU(2|2)/U(1)^2$ whose bosonic part is Euclidean $AdS^2 \times S^2.$
In this case there are varieties to scale grassmannian coordinates.
We first give an example of a similar scaling to the two-dimensional
case. In this case, we can easily perform the same calculation as the 
supersphere cases and obtain covariant derivatives which commute
with supercharges. Among eight supersymmetries,
four of them are dynamical and generate space-time supersymmetries.
Space-time translations in only two directions out of four 
appear in the anticommutators of supersymmetries and after imposing
chiral constraints this system becomes essentially two-dimensional.
We also give another example where all four space-time 
translation generators appear in the anticommutators.
In the latter half of this section, we explain briefly that
the Seiberg's $d=4$ noncommutative superspace \cite{Seiberg}
can be understood as a constrained system. 

\par
In the appendix A, we review a general method using the Cartan one-forms
to obtain mutually commutative sets of generators, left and right
multiplications on the group manifold. They become covariant derivatives
and guiding center coordinates in our cases.
We then explain a method for the generalized 
In$\ddot{\rm o}$n$\ddot{\rm u}$-Wigner contractions.
In the following appendices,
we give some detailed calculations for 
supercovariant derivatives and global charges (guiding center 
coordinates).

\section{Fuzzy sphere and flat limit}
In this section, we review
the bosonic fuzzy sphere as a warming up.
A fuzzy sphere and field theories on it 
can be constructed by matrix models based on the $su(2)$ Lie algebra.
(For more details, see \cite{alek,fuzzy2,IKTW}, for example.)
Take representation matrices $\hat{l}_i$ 
of $su(2)$ with an angular momentum
$L$. The size of the matrices is $N=2L+1$. 
The noncommutative coordinates are defined as 
$\hat{x}_i=\alpha \hat{l}_i$. Then the radius of the sphere $r$
is $r^2=\alpha^2 L(L+1)$. Any $N \times N$ matrix can be
expanded in terms of noncommutative spherical harmonics $\hat{Y}_{lm}$
where $l$ runs from $0$ to $2L$.
The coordinates satisfy the noncommutative algebra;
\begin{equation}
[\hat{x}_i, \hat{x}_j] = i \alpha \epsilon_{ijk} \hat{x}_k. 
\label{fuzzyalg}
\end{equation}
\par
We now take the noncommutative flat limit of the fuzzy sphere
algebra (\ref{fuzzyalg}).
This corresponds to considering the vicinity of the north pole 
and scaling $(x_1,x_2)$ coordinates  so that
the noncommutativity is fixed.
This can be achieved by redefining coordinates as 
\begin{equation}
\hat{x}= \sqrt{\frac{\Theta}{r}} \hat{x}_1, 
\ \  \ \hat{y}= \sqrt{\frac{\Theta}{r}} \hat{x}_2.
\label{bosonscale}
\end{equation}
They satisfy the commutation relation 
$ [\hat{x}, \hat{y}] = i \Theta$.
$\Theta$ is a noncommutative parameter.
\subsection{Fuzzy sphere as the lowest Landau level}
The noncommutative coordinates on the fuzzy sphere 
can be understood as 
the guiding center coordinates on an ordinary sphere in a
magnetic monopole at the origin. Let us see this explicitly.
We consider a particle moving on a sphere with radius $r$
in the field of a monopole put at the origin. 
The Hamiltonian is given by
\begin{equation}
H = \frac{1}{2m}D_i^2
\end{equation}
where $D_i$ is the covariant derivative 
$D_i=-i(\partial_i - i eA_i$).
In a monopole magnetic field, the commutator becomes
\begin{equation}
[D_i, D_j] =  -i e g \epsilon_{ijk} \frac{x_k}{r^3}.
\end{equation}
The Hamiltonian is written in terms of the following deformed
$su(2)$ generators  \cite{Coleman}
\begin{equation}
K_i = \epsilon_{ijk} x_j D_k + eg \frac{x_i}{r}
\end{equation}
as 
\begin{equation}
H = \frac{K_i^2 -e^2 g^2}{2 m r^2}.
\end{equation}
A consistency condition requires that the allowed values
of the total angular momentum  generated by $K_i$ are
\begin{equation}
L=|eg|, |eg|+1, \cdots.
\end{equation}
Each state with a fixed $L$ corresponds to each Landau level state. 
The lowest Landau level states have degeneracy $2|eg|+1$.
Such degeneracies are described by the eigenvalue of $K_3$
and they are generated by acting the 
$K_{-}$ operator  on the highest weight state. 
To describe the degeneracy in each Landau level,
we can define the guiding center coordinates $X_i$ as
\begin{equation}
\hat{X}_i =  \alpha K_i
\end{equation}
where $\alpha$ is defined so that the radius of 
$\hat{X}_i$ becomes $r$.
Then $r^2 =\alpha^2 L(L+1)$. 
Since they commute with the Hamiltonian, they are constants
of motion and can be interpreted as the guiding center 
coordinates of the cyclotron motion on the sphere. 
In the large $g$ limit, 
$\alpha = r / eg$ and
the first term in $K_i$ can be neglected compared to the 
second term. Then $\hat{X}_i$ is identified with the 
commutative coordinate $x_i$. 
This means that the cyclotron 
radius becomes small in the large magnetic field limit.
These guiding center coordinates are noncommutative;
\begin{equation}
[\hat{X}_i, \hat{X}_j] = i \epsilon_{ijk} \alpha \hat{X}_k.
\end{equation}
This is nothing but the commutation relation 
(\ref{fuzzyalg}) on the fuzzy sphere.

\subsection{Flat limit of fuzzy sphere}
In this subsection, we apply the general method (see Appendix A) 
to obtain covariant derivatives and guiding center coordinates to the
simplest case.
We first parametrize the group manifold of $SU(2)$ by 
$g=exp(i{\bf L}\cdot{\bf x})$. Following the method of
the (generalized) In$\ddot{\rm o}$n$\ddot{\rm u}$-Wigner contractions, 
we can obtain a flat limit by taking the following scaling
of the parameter space.
We first take the scaling 
\begin{equation}
 x \rightarrow s x, \ y \rightarrow s y, \
  z \rightarrow   z.
\end{equation}
Then if we take up to the second order of $s^2$, we can 
obtain the algebra
\bea
&&\left[L_{x[1]},L_{y[1]}\right]=i{L}_{z[2]}~,~
\left[L_{x[1]},{L}_{z[2]}\right]=\left[L_{y[1]},{L}_{z[2]}\right]=0~\nn\\
&&\left[L_{z[0]},{L}_{x[1]}\right]=iL_{y[1]}~,
~\left[L_{y[1]},{L}_{z[0]}\right]=iL_{x[1]}.
~~\label{flatalg}
\eea
The generator $L_{z[2]}$ is a center and can be considered as
a constant. The generator $L_{z[0]}$ is a rotation generator
on the two dimensional plane. 
If we instead take the scaling of $z$ as 
\begin{equation}
z \rightarrow s^2 z,
\end{equation}
the generator $L_{z[0]}$ disappears and the algebra of the
noncommutative plane which is given by the first 
line of (\ref{flatalg}) can be obtained. 
In the following we 
consider this case.
\par
The covariant derivatives(right multiplication generators)
 of the algebra generated by
$L_{x[1]}, L_{y[1]}$ and the center $L_{z[2]}$ 
can be constructed 
by the Cartan 1-form $g^{-1}dg=dx^me_m{}^aT_a$ 
as ${\bf D}_a= -i (e^{-1})_a{}^m \partial_m$.
We take the parametrization 
$(x,y,\phi)$ for $T_a=(L_{x[1]},L_{y[1]},L_{z[2]})$.
Then the covariant derivatives are given as 
\bea
{\bf D}_x&=&\frac{\partial}{i\partial x}+\frac{y}{2}\frac{\partial}{i\partial \phi}\nn\\
{\bf D}_y&=&\frac{\partial}{i\partial y}-\frac{x}{2}\frac{\partial}{i\partial \phi}\label{ccddvv}~~~\\
{\bf D}_z&=&\frac{\partial}{i\partial \phi}.
\eea
They satisfy the algebra
\bea
\left[{\bf D}_x, {\bf D}_y\right]=i{\bf D}_z~~~.\label{DDz}
\eea

The global charges (left multiplication generators)
are constructed similarly and given by
\bea
\hat{\bf K}_x&=&\frac{\partial}{i\partial x}
-\frac{y}{2i}\frac{\partial}{\partial \phi}\nn\\
\hat{\bf K}_y&=&\frac{\partial}{i\partial y}+\frac{x}{2i}\frac{\partial}{\partial \phi}
\nn\\
\hat{\bf K}_z&=&\frac{\partial}{i\partial \phi}   \label{chaxyz}~~~.
\eea
These global charges commute with the covariant derivatives.
\par
Since $\phi$ is a coordinate conjugate to the center $L_{z[2]}$
and $\partial/\partial \phi$ is a generator to multiply a constant
$(L_{z[2]})$ on $g$,
we can fix it 
as a constant: ${\bf D}_z=\frac{\partial}{i\partial \phi}=-\Theta^{-1}\neq 0$. 
Then the commutator becomes
\bea
\left[{\bf D}_x, {\bf D}_y\right]=-i\Theta^{-1}.\label{DDT}
\eea
We  define the noncommutative coordinates $X,Y$
(the guiding center coordinates) as
\bea
X&=&\frac{1}{2}\left(x-2\Theta  \frac{\partial}{i\partial y}\right)=-\Theta \hat{\bf K}_y=x-\Theta {\bf D}_y\nn\\
Y&=&\frac{1}{2}\left(y+2\Theta  \frac{\partial}{i\partial x}\right)=\Theta \hat{\bf K}_x =y+\Theta {\bf D}_x
\label{XYxy}~~~.
\eea
They satisfy the algebra
\bea
\left[X,Y\right]={i\Theta}~~~.\label{ncc}
\eea
The transformation from $(x,y, p_x, p_y)$ to $(X,Y,{\bf D}_x, {\bf D}_y)$
is  familiar in the two dimensional system in a 
constant magnetic field.
If we consider a particle constrained in the lowest Landau level,
we impose the lowest Landau level condition (for $\Theta >0$)
\begin{equation}
( {\bf D}_x - i {\bf D}_y) |LLL \rangle =0.
\end{equation}
This constraint can be made as the second class by imposing 
${\bf D}_x=0,~{\bf D}_y=0$.
The Dirac bracket for these second class constraints is given by
\bea
\left[{\cal O}_1,{\cal O}_2\right]_D
=\left[{\cal O}_1,{\cal O}_2\right]
+i\Theta\left[{\cal O}_1,{\bf D}_x\right]\left[{\bf D}_y,{\cal O}_2\right]
-i\Theta\left[{\cal O}_1,{\bf D}_y\right]\left[{\bf D}_x,{\cal O}_2\right]~~~.
\eea
The Dirac bracket of the original coordinates becomes
\bea
\left[x,y\right]_D=i\Theta~~~.
\eea
On the other hand using facts that $(X,Y)$ are equal to $(x,y)$ up to 
the second class constraints
and they commute with $({\bf D}_x, {\bf D}_y)$,
we have
\bea
\left[x,y\right]_D=\left[X,Y\right]_D=\left[X,Y\right]=i\Theta~~~.
\eea
Therefore the Dirac bracket of the original coordinates gives
the noncommutative coordinate algebra \bref{ncc}. 
In this picture, the noncommutative space is considered as a
space whose phase space degrees of freedom is reduced by 
the constraint. 
\par

\vskip 6mm
\section{Noncommutative flat superspace from osp(1$\mid$2)}\par
In the following sections we 
generalize the method in the previous section to superspaces.
A construction of noncommutative space 
is performed by generalization of 
ordinary Lie algebras to super Lie algebras. 
In this way, we can systematically construct supermatrix models
and field theories on noncommutative homogeneous superspaces.
This was first studied in \cite{Grosse} for the case of scalar
multiplets with $osp(1|2)$ symmetry. A gauge theory can be 
similarly constructed \cite{HIU22}.
We then take the 
flat limit with an appropriate scaling of operators.
We can also understand the noncommutative superspace as a 
constrained system whose phase space dimension is reduced by
the second class constraints.
This give an interpretation that the noncommutative superspace
is a supersymmetric analog of the lowest Landau level system.
In this section, we take  the $osp(1|2)$ super Lie
algebra. This gives a two dimensional noncommutative supersphere
with two real supercharges on the fuzzy sphere. 
\subsection{osp(1$\mid$2) algebra and fuzzy supersphere}
The graded commutation relations of $osp(1|2)$ algebra are given by 
\begin{eqnarray}
\left[\hat{l}_i, \hat{l}_j\right] &=& i\epsilon_{ijk}\hat{l}_k, \nonumber \\
\left[\hat{l}_i, \hat{v}_\alpha\right] &=&
  \frac{1}{2}\left(\sigma_i\right)_{\beta\alpha}\hat{v}_\beta, \\
\label{ospalgebra}
\left\{\hat{v}_\alpha, \hat{v}_\beta\right\} &=&
  \frac{1}{2}\left(C\sigma_i\right)_{\alpha\beta}\hat{l}_i, \nonumber
\end{eqnarray}
where $C=i\sigma_2$. The even part of this algebra is $su(2)$ which 
is generated by $\hat{l}_i \ (i=1, 2, 3)$ and the odd generators 
$\hat{v}_\alpha \ (\alpha=1, 2)$ are $su(2)$ spinors.
In this paper, we also write $\hat{v}_1=\hat{v}_+$ and $\hat{v}_2=\hat{v}_-$.
The irreducible representations of $osp(1|2)$
algebra~\cite{representation} are characterized
by the values of the Casimir operator 
$\hat{K}_2=\hat{l}_i\hat{l}_i
+C_{\alpha\beta}\hat{v}_\alpha \hat{v}_\beta=L(L+\frac{1}{2})$ 
where quantum number $L$ is 
called super spin and 
$L\in\mathbb{Z}_{\geq 0}/2$. 
Each representation consists of spin $L$ and $L-\frac{1}{2}$
representations of $su(2)$, $|L, l_3\rangle, |L-\frac{1}{2}, l_3\rangle$
and its dimension is $N=(2L+1)+2L=4L+1$.
\par
The condition $\hat{K}_2=L(L+\frac{1}{2})$ 
defines the two-dimensional super sphere.
Consider polynomials $\Phi(l_i, v_\alpha)$ of the representation
matrices $l_i$ and $v_\alpha$ with super spin $L$. 
Let us denote the space spanned by $\Phi(l_i, v_\alpha)$ as ${\cal A}_L$.
The $osp(1|2)$ algebra acts on ${\cal A}_L$ by the three kinds of action, 
the left action $(\hat{l}_i^L, \hat{v}_\alpha^L)$, the right action 
$(\hat{l}_i^R, \hat{v}_\alpha^R)$ and the adjoint action
$(\hat{{\cal L}}_i\equiv \hat{l}_i^L-\hat{l}_i^R, \hat{{\cal V}}_\alpha=\hat{v}_\alpha^L-\hat{v}_\alpha^R)$,
\begin{eqnarray}
\hat{l}_i^L \Phi = l_i \Phi, && \hat{v}_\alpha^L \Phi = v_\alpha \Phi, \\
\hat{l}_i^R \Phi = \Phi l_i, && \hat{v}_\alpha^R \Phi = \Phi  v_\alpha, \\
\hat{{\cal L}}_i \Phi = [l_i, \Phi], && 
  \hat{{\cal Q}}_\alpha \Phi = [v_\alpha, \Phi].
\end{eqnarray}
We can define supersymmetrized spherical harmonics
$\hat{Y}_{km}^S$ which are generalization of the ordinary
spherical harmonics to the super sphere (see \cite{Grosse2}
for the details). $k$ can take either an integer or 
half an integer value and they are bosonic or fermionic
functions respectively.
Any $N \times N$ supermatrix can be expanded in terms of
the super spherical harmonics as
\begin{eqnarray}
\Phi(l_i, v_\alpha)=\sum_{k=0, 1/2, 1, \cdots}^{2L} \phi_{km}\hat{Y}^S_{km},
\end{eqnarray} 
where the coefficient $\phi_{km}$ for the even (odd) spherical harmonics
is Grassmann even (odd). 
We can map the supermatrix $\Phi(l_i, v_\alpha)$ to a function 
on the superspace $(x_i, \theta_\alpha)$ by 
\begin{equation}
 \Phi(l_i, v_\alpha)\longrightarrow 
  \phi(x_i, \theta_\alpha)=\sum_{k,m} \phi_{km}y^S_{km}(x_i, \theta_\alpha),
\end{equation}
where $y^S_{km}(x_i, \theta_\alpha)$ are ordinary superspherical
functions.
A product of supermatrices is mapped to a noncommutative star product
of functions.
An explicit form of the star product is given in \cite{Balachandran}.
\par
In addition to the $osp(1|2)$ generators $(\hat{l}_i, \hat{v}_{\alpha})$,
we can define additional generators with which they form bigger
algebra $osp(2|2)$.
These additional generators are 
\begin{eqnarray}
&& \hat{\gamma} = -\frac{1}{L+1/4}
(C_{\alpha \beta}\hat{v}_{\alpha} \hat{v}_{\beta} + 2 L(L+{1 \over 2})) \\
\label{addg}
&& \hat{d}_{\alpha} = [\hat{\gamma}, \hat{v}_{\alpha}]=\frac{1}{2(L+1/4)}
(\sigma_i)_{\beta \alpha} (\hat{v}_{\beta} \hat{l}_i + 
\hat{l}_i \hat{v}_{\beta}).
\label{addd}
\end{eqnarray}
Commutation relations for the additional generators are given by
\begin{eqnarray}
\left[\hat{\gamma}, \hat{v}_\alpha\right] &=& \hat{d}_\alpha, \nonumber \\
\left[\hat{\gamma}, \hat{d}_\alpha\right] &=& \hat{v}_\alpha,  \n
\left[\hat{\gamma}, \hat{l}_i\right] &=& 0, \nonumber \\
\left[\hat{l}_i, \hat{d}_\alpha\right] &=&
  \frac{1}{2}\left(\sigma_i\right)_{\beta\alpha}\hat{d}_\beta,  \\ 
\left\{\hat{d}_\alpha, \hat{d}_\beta\right\} &=&
  -\frac{1}{2}\left(C\sigma_i\right)_{\alpha\beta}\hat{l}_i, \nonumber \\
\left\{\hat{v}_\alpha, \hat{d}_\beta\right\} &=&
  -\frac{1}{4}C_{\alpha\beta}\hat{\gamma}. \nonumber 
\end{eqnarray}
The adjoint action of 
the fermionic generators $D_{\alpha}=\mbox{adj}~\hat{d}_\alpha$ 
plays a role of the covariant derivatives. 
On the other hand, the adjoint action 
of the original fermionic generators 
$Q_{\alpha}=\mbox{adj}~\hat{v}_\alpha$ 
are interpreted as super symmetry generators. We will show that they
commute in the flat limit. These additional generators also
play an important role in constructing kinetic terms
for a scalar multiplet on the super sphere \cite{Grosse}.
\par
The commutative limit is  discussed in \cite{Grosse}
and the fuzzy super sphere becomes the ordinary
two-dimensional supersphere with two real grassmannian
coordinates. 
This limit can be taken by keeping the radius $r$
of the sphere fixed and taking the large $L$ limit. 
\subsection{Flat noncommutative superspace}
We now take a flat limit. Namely we consider the vicinity of
the north pole. For the bosonic generators, we perform the
same scaling as (\ref{bosonscale}) to obtain the flat noncommutative
coordinates.
For the fermionic generators, in order to keep the noncommutativity,
one possible choice is to scale both of $v_{\alpha}$ as $\sqrt{L}$.
Then from the $osp(1|2)$ algebra (\ref{ospalgebra}) we can read that
the anticommutator of $v_+$ and $v_-$ survives to be a constant.
 The other anticommutators among $v$ vanish.
But with this choice we can easily see that the supersymmetry
algebras acting on the noncommutative superspace
become trivial, i.e. the anticommutators between 
supercharges do not generate translation of the space.
All supersymmetries become non-dynamical symmetries.
So we need to take another scaling limit
where $v_+$ and $v_-$ are scaled asymmetrically.
We define new superspace coordinates as
\begin{eqnarray}
\hat{x}_i &=& \sqrt{\frac{\Theta}{L}} \hat{l}_i \ \mbox{    for $i=1,2$}  \n
\hat{\theta}_+ &=& \sqrt{2} \left( \frac{\Theta}{L} \right)^{1/4} \hat{v}_+  \\
\label{ncscaling1}
\hat{\theta}_- &=& \sqrt{2} \left( \frac{\Theta}{L} \right)^{3/4} \hat{v}_- . 
\nonumber
\end{eqnarray}
We multiply a constant on $\hat{\theta}_{\alpha}$ for convenience.
Since we are considering around the north pole, 
the $\hat{l}_3$ is scaled as $L$.
With the above scaling, the algebra among the coordinates become
\begin{eqnarray}
&& [\hat{x}_+,\hat{x}_-]= 2 \Theta \hspace{6mm} \mbox{or} \hspace{6mm}
[\hat{x}_1,\hat{x}_2]=i \Theta \n 
&&\{ \hat{\theta}_+, \hat{\theta}_+\} =  \hat{x}_+ \n
&& \{ \hat{\theta}_+, \hat{\theta}_-\} =  - \Theta \\
&& [\hat{x}_-, \hat{\theta}_+] = \hat{\theta}_- . \nonumber
\end{eqnarray}
where we have defined $\hat{x}_{\pm}= \hat{x}_1 \pm i \hat{x}_2$. 
All the other commutators vanish. 
By redefining the coordinates as
\begin{eqnarray}
&& \hat{\varphi}_+ = \hat{\theta}_+ + \frac{1}{2 \Theta} 
\hat{x}_+ \hat{\theta}_-  \\
&& \hat{\varphi}_- = \hat{\theta}_-,
\end{eqnarray}
the noncommutativity of the coordinates is written simply as a 
canonical form;
\begin{equation}
\label{com-phi}
[\hat{x}_+,\hat{x}_-]= 2 \Theta, \ \ \ 
\{ \hat{\varphi}_+, \hat{\varphi}_-\} =  - \Theta.
\end{equation}
\par
The scaling of 
the additional generators $\hat{d}_{\alpha}$
can be automatically determined from the scaling (\ref{ncscaling1})
because they are written in terms of the $osp(1|2)$ generators.
The scaled generators are defined by
\begin{eqnarray}
&& \hat{d}_+' = \sqrt{2} \left( \frac{\Theta}{L} \right)^{-1/4} \hat{d}_+
  = \hat{\theta}_+ + \frac{1}{\Theta} \hat{\theta}_- \hat{x}_+
= \hat{\varphi}_+ + \frac{1}{2 \Theta}\hat{\varphi}_- \hat{x}_ +, \\
&& \hat{d}_-' = \sqrt{2} \left( \frac{\Theta}{L} \right)^{-3/4} \hat{d}_-
= - \hat{\theta}_- = -\hat{\varphi}_-.
\end{eqnarray}
The anticommutator with $\hat{\theta}$ is given by
\begin{equation}
\{  \hat{d}_{\alpha}', \hat{\theta}_{\beta}\} = - 
C_{\alpha \beta} \Theta,
\end{equation}
which is consistent with the scaling of $\hat{\gamma}$.
$\hat{\gamma}$ becomes a constant and commutes with all the other 
generators.
\par
Now we define generators of supersymmetry transformations and 
covariant derivatives which mutually commute by
\begin{eqnarray}
&& P_{\pm} = \pm {1 \over 2} \mbox{ adj } \hat{x}_{\pm} \\
&& Q_{\alpha} = \mbox{ adj }\hat{\theta}_{\alpha}.
\end{eqnarray}
They generate  supersymmetry transformations
on the superspace coordinates $(\hat{x}, \hat{\varphi})$. 
They satisfy the  supersymmetry algebra
\begin{eqnarray}
&& [P_+,P_-]=0 \\
&& [P_-, Q_+]= - {1 \over 2} Q_- \\
&& \{ Q_+, Q_-\} =0 \\
&& \{ Q_+, Q_+\} =  2 P_+ \\
&& \{ Q_-, Q_-\} =0 .
\end{eqnarray}
$Q_+$ is a dynamical supersymmetry and generates space-time
translation into $x_-$ direction. But the $Q_-$ is a non-dynamical
supersymmetry and its anticommutator vanishes. This is caused by
the asymmetric scaling of the coordinates. Because of this
asymmetric scaling, we cannot take a further limit to obtain an 
ordinary two-dimensional superspace with two dynamical supersymmetries.
\par 
Covariant derivatives can be defined similarly 
by the adjoint action of $\hat{d}_{\alpha}'$;
\begin{equation}
D_{\alpha} = \mbox{ adj } \hat{d}_{\alpha}'.
\end{equation}
They anticommute with $Q_{\alpha}$; $\{ D_{\alpha}, Q_{\beta}\}=0$.
Their commutation relations are
\begin{eqnarray}
&& \{ D_+, D_+\} = - 2 P_+ \\
&& \{ D_+,D_-\}=\{D_-,D_-\}=0 \\
&& [P_-, D_+]= -  \frac{1}{2} D_-
\end{eqnarray}
\par
Functions on the superspace $(\hat{x}, \hat{\varphi})$ are given as  
$N \times N$ supermatrices.
Generically they depend on full set of supercoordinates:
$\Phi(\hat{x}_+, \hat{x}_-, \hat{\varphi}_+, \hat{\varphi}_-)$.
We can consistently 
constrain them by imposing the chiral constraint as
\begin{equation}
  D_- \Phi = - [\hat{\varphi}_-, \Phi]=0.
\end{equation}
This automatically leads to
\begin{equation}
  Q_- \Phi =0
\end{equation}
and the superfield $\Phi$ depends only on 
$(\hat{x}_+, \hat{x}_-, \hat{\varphi}_-)$.
This is the chiral superfield and supersymmetry is generated
by $Q_+$ whose anticommutator becomes $P_+$.
To summarize, on the chiral superfields 
$\Phi(\hat{x}_+,\hat{x}_-,\hat{\varphi}_-)$, the algebra 
of supersymmetry $Q_+$, translations $P_{\pm}$ and the
covariant derivative $D_+$ is given by
\begin{eqnarray}
&& \{ Q_+,Q_+\}=-\{ D_+, D_+\}=2 P_+ \\
&& [P_{\pm}, Q_+]=[P_{\pm}, D_+]=0 \\
&& [P_+, P_-] =0.
\end{eqnarray}
They are written  as differential operators on chiral superfields 
$\Phi(\hat{x}_+, \hat{x}_-, \hat{\varphi}_-)$;
\begin{eqnarray}
&& P_{\pm} = \Theta \frac{\partial}{\partial \hat{x}_{\mp}} \\
&& Q_+ = - \Theta \frac{\partial}{\partial \hat{\varphi}_-} 
          - \hat{\varphi}_- \frac{\partial}{\partial \hat{x}_{-}} \\
&& D_+ = - \Theta \frac{\partial}{\partial \hat{\varphi}_-} 
          + \hat{\varphi}_- \frac{\partial}{\partial \hat{x}_{-}}
\end{eqnarray} 
We can further constrain the superfield by $P_-$:
\begin{equation}
  P_- \Phi =0.
\end{equation}
Then the superfield becomes independent of the $\hat{x}_+$ coordinate:
$\Phi(\hat{x}_-,\hat{\phi}_-)$,
and the system becomes essentially one-dimensional.
\subsection{Magnetic field in 2d superspace}\par
The noncommutative superspace can be understood as a system
restricted by some constraints analogous to the lowest
Landau level states in the bosonic case. 
In this subsection, we first derive mutually commutative
set of differential operators
acting on superparticles in the commutative superspace.
They correspond to the covariant derivatives in a magnetic field
and the guiding center coordinates discussed in the previous section.
\par
In order to obtain mutually commutative set of generators
on the flat space, we again begin with the $osp(1|2)$ algebra 
(\ref{ospalgebra}), where we denote the generators as 
$\hat{l}_i=L_i, \sqrt{2}\hat{v}_\alpha=Q_\alpha$,
and take the In$\ddot{\rm o}$n$\ddot{\rm u}$-Wigner contraction. 
We parametrize the group manifold
of $osp(1|2)$ by $(x,y,\phi,\theta_{\pm})$ as 
$g=exp(i {\bf x}\cdot{\bf L}) exp (i\theta^{\alpha} Q_{\alpha})$.
We then take the scaling as
\bea
x, y\to sx, sy~,~
\phi\to  s^2 \phi~ ,~
\theta^+\to \sqrt{s} \theta^+~,~
\theta^-\to \sqrt{s}^3 \theta^-
\label{IWsp}
\eea
and take $s\to 0$ limit. 
\par
If we take up to the second order of $s$, we have the algebra
\bea
\left[L_{x[1]},L_{y[1]}\right]&=&iL_{z[2]}\nn\\
\left[L_{-[1]},Q_{+[1/2]}\right]&=&Q_{-[3/2]} \nn\\
\left\{Q_{+[1/2]},Q_{+[1/2]}\right\}&=&L_{+[1]}\nn\\
\left\{Q_{+[1/2]},Q_{-[3/2]}\right\}&=&-L_{z[2]}~~~~~,~~{\rm others}=0~~.
\label{ncst}
\eea 

Covariant derivatives are defined as the right multiplication on 
the group manifold generated by this algebra.
They are calculated in the appendix
resulting as
\bea
{\bf D}_x&=&\frac{\partial}{i\partial x}
+\frac{y}{2i}\frac{\partial}{\partial \phi}
-\frac{1}{2}\theta^+\frac{\partial}{\partial \theta^-}\nn\\
{\bf D}_y&=&\frac{\partial}{i\partial y}-\frac{x}{2i}\frac{\partial}{\partial \phi}
+\frac{1}{2i}\theta^+\frac{\partial}{\partial \theta^-}\nn\\
{\bf D}_z&=&\frac{\partial}{i\partial \phi}   \label{covosp} \\
D_+&=&\frac{\partial}{i\partial \theta^+}
-\frac{1}{2}\theta^-\frac{\partial}{\partial \phi}
+\frac{x+iy}{4i}\theta^+\frac{\partial}{\partial \phi}
+\frac{1}{2}\theta^+(\frac{\partial}{\partial x}+i
\frac{\partial}{\partial y})\nn\\
D_-&=&\frac{\partial}{i\partial \theta^-}
-\frac{1}{2}\theta^+\frac{\partial}{\partial \phi} \nn
\eea
satisfying
\bea
\left[{\bf D}_x,{\bf D}_y\right]&=&i{\bf D}_z\nn\\
\left[{\bf D}_-,D_+\right]&=&D_- \nn\\
\left\{D_+,D_+\right\}&=&{\bf D}_+\nn\\
\left\{D_+,D_-\right\}&=&-{\bf D}_z~~~,~~{\rm others}=0~~.
\label{covosp2}
\eea 
These commutators are supersymmetric generalization of
(\ref{DDz}). The terms in the r.h.s.  with a ${\bf D}_z$ term
are interpreted as the effect of the background magnetic field.
\par
The global charges (left multiplications) 
are similarly calculated in the appendix resulting as
\bea
\hat{\bf L}_x&=&\frac{\partial}{i\partial x}
-\frac{y}{2i}\frac{\partial}{\partial \phi}\nn\\
\hat{\bf L}_y&=&\frac{\partial}{i\partial y}+\frac{x}{2i}\frac{\partial}{\partial \phi}
\nn\\
\hat{\bf L}_z&=&\frac{\partial}{i\partial \phi}   \label{glocha}\\
\hat{Q}_+&=&\frac{\partial}{i\partial \theta^+}
+\frac{1}{2}\theta^-\frac{\partial}{\partial \phi}
-\frac{x+iy}{2}\frac{\partial}{\partial \theta^-}
-\frac{1}{2}\theta^+(\frac{\partial}{\partial x}+i
\frac{\partial}{\partial y})\nn\\
\hat{Q}_-&=&\frac{\partial}{i\partial \theta^-}
+\frac{1}{2}\theta^+\frac{\partial}{\partial \phi}\nn
\eea
satisfying
\bea
\left[\hat{\bf L}_x,\hat{\bf L}_y\right]&=&-i\hat{\bf L}_z\nn\\
\left[\hat{\bf L}_-,\hat{Q}_+\right]&=&-\hat{Q}_- \nn\\
\left\{\hat{Q}_+,\hat{Q}_+\right\}&=&-\hat{\bf L}_+\nn\\
\left\{\hat{Q}_+,\hat{Q}_-\right\}&=&\hat{\bf L}_z~~~,~~{\rm others}=0~~.
\eea 
These global charges commute/anticommute with covariant derivatives
including $\left\{ Q, D\right\}=0$. These operators are interpreted 
as the guiding center coordinates in a constant magnetic field
on the two dimensional superspace.
\par
The covariant derivatives (\ref{covosp}) can be considered 
as the supercovariant derivatives (\ref{1.1}), (\ref{1.2})
in a constant magnetic field.
In (\ref{covosp}), 
the terms containing derivatives with respect to $\phi$
are contributions from the gauge fields.
Applying the definition of the field strength given 
in the introduction (\ref{superstrength}), 
the above system has  a constant magnetic
field in $(x,y)$ and $(\theta_+, \theta_-)$ directions;
\begin{equation}
F_{x,y}= B, \ \ \  F_{\theta_+, \theta_-}=iB
\end{equation}
where we write ${\bf D}_z=B$. 
In the next subsection we show that this induces the noncommutativity
on superspace.
\subsection{Noncommutative superspace as the lowest Landau level system}
We now calculate the Dirac bracket of the super coordinates
by imposing the 'lowest Landau level' constrains.
Since $\phi$ is a coordinate conjugate to the 
center $L_{z[2]}$ we can fix it as a constant;
${\bf D}_z=-\Theta^{-1}\neq 0$.
This leads to the 
fermionic noncommutative algebra
as well as the bosonic noncommutative algebra \bref{DDT},
\bea
\left[{\bf D}_x,{\bf D}_y\right]&=&-i\Theta^{-1}~~,~~
\left\{D_+,D_-\right\}=\Theta^{-1}~~.
\eea
These algebras are shown to induce  noncommutative fermionic
coordinates as well as noncommutative bosonic coordinates;
\bea
\left[X,Y\right]=i\Theta~~,~~\left\{\psi^+,\psi^-\right\}=\Theta~~~.
\label{XYpsi}
\eea
\par
The 'lowest Landau level' conditions are 
\begin{eqnarray}
&&  ({\bf D}_x - i {\bf D}_y) |LLL \rangle =0, \\
&&  D_- |LLL \rangle =0.
\end{eqnarray}
Analogous to the bosonic case the canonical analysis can be performed.
Second class constraints are given by
${\bf D}_x={\bf D}_y=D_+=D_-=0$ and
the Dirac bracket for the system is calculated as 
(we have dropped terms up to the second class constraints)
\bea
&&\hspace*{-7mm}
\left[{\cal O}_1,{\cal O}_2\right\}_D
=\left[{\cal O}_1,{\cal O}_2\right\}\label{Dncsp}\\
&&+\Theta\left(
i\left[{\cal O}_1,{\bf D}_x\right\}\left[{\bf D}_y,{\cal O}_2\right\}
-i\left[{\cal O}_1,{\bf D}_y\right\}\left[{\bf D}_x,{\cal O}_2\right\}
-\left[{\cal O}_1,{D}_+\right\}\left[{D}_-,{\cal O}_2\right\}
-\left[{\cal O}_1,{D}_-\right\}\left[{D}_+,{\cal O}_2\right\}
\right).\nn
\eea
Then the Dirac brackets of the original coordinates become
\bea
\left[x,y\right]_D=i\Theta~,~
\{\theta^+,\theta^-\}_D=\Theta~,~
{\rm others}=0.
\eea
We can introduce noncommutative guiding center coordinates 
which are written in terms of the global charges (${\bf L}_i, Q_{\alpha}$)
in such a way that
they are equal to the original coordinates up to 
the second class constraints.
The results are
\bea
X&=&\frac{1}{2}\left(x-2\Theta  \frac{\partial}{i\partial y}\right)
=-\Theta \hat{\bf L}_y=x-\Theta {\bf D}_y-\frac{i}{2}\Theta \theta^+D_-\nn\\
Y&=&\frac{1}{2}\left(y+2\Theta  \frac{\partial}{i\partial x}\right)
=\Theta \hat{\bf L}_x =y+\Theta {\bf D}_x+\frac{i}{2}\Theta \theta^+D_-\nn\\
\psi^+&=&\frac{1}{2}\theta^++\Theta\frac{\partial}{\partial \theta^-}
=i\Theta \hat{Q}_-
=\theta^++i\Theta D_-\nn\\
\psi^-&=&\frac{1}{2}\theta^-+\Theta\frac{\partial}{\partial \theta^+}
=i\Theta \hat{Q}_+-\frac{\Theta}{2}\psi^+\hat{\bf L}_+
=\theta^-+i\Theta D_++\frac{\Theta}{2}{\bf D}_+\theta^+
\label{XYxypsi}~~~.
\eea 
They satisfy
\bea
\left[X,Y\right]&=&\left[x,y\right]_D=i\Theta~\nn\\
\{\psi^+,\psi^-\}&=&\{\theta^+,\theta^-\}_D
=\Theta~~.
\eea
and all the other commutators vanish.
This algebra is nothing but the canonical commutation relation 
(\ref{com-phi}) 
for the redefined coordinates.
The fermionic coordinates $\psi^{\alpha}$ are related to $\hat{\varphi}$ as
\begin{equation}
 \hat{\varphi}_{\alpha}= \epsilon_{\alpha \beta} \psi^{\beta}.
\end{equation}

\indent
\section{Noncommutative superspace from su(2$\mid$1)}
In this section, we consider noncommutative superspace
based on $su(2|1)$ super algebra (or equivalently $osp(2|2)$ algebra).
This gives two-dimensional supersphere with four real supercharges.
This type of noncommutative superspace was studied in \cite{Klimcik}
though an explicit relation to our case is not manifest.
\par
The graded commutation relations of $su(2|1)$ algebra are given by
\begin{eqnarray}
 \begin{array}{ll}
  \displaystyle{\left[\hat{l}_i, \hat{l}_j\right] 
   = i \epsilon_{ijk}\hat{l}_k,} & \\
  \displaystyle{\left[\hat{l}_i, \hat{q}_\alpha\right]
   = -\frac{1}{2}{(\sigma_i)_\alpha}^\beta \hat{q}_\beta,} &
   \displaystyle{\left[\hat{l}_i, \hat{\bar{q}^\alpha}\right]
   =\frac{1}{2}{(\sigma_i)_\beta}^\alpha\hat{\bar{q}^\beta},} \\
  \displaystyle{\left[\hat{B}, \hat{q}_\alpha\right] 
   = \frac{1}{2}\hat{q}_\alpha,} &
   \displaystyle{\left[\hat{B}, \hat{\bar{q}^\alpha}\right] 
   = -\frac{1}{2}\hat{\bar{q}^\alpha},} \\
  \displaystyle{\left\{\hat{q}_\alpha, \hat{\bar{q}^\beta}\right\} 
   = {(\sigma_i)_\alpha}^\beta \hat{l}_i 
   + \delta_\alpha^\beta \hat{B},} & 
   \mbox{others} = 0.
 \end{array}
\end{eqnarray}
This contains an $osp(1|2)$ super algebra as a subalgebra.
There are two Casimir operators.
The second Casimir operator of $su(2|1)$ algebra is given by 
\begin{equation}
 \hat{K}_2 = \hat{l}_i\hat{l}_i-\hat{B}^2
  +\frac{1}{2}\left(\hat{q}_\alpha\hat{\bar{q}^\alpha}
	      -\hat{\bar{q}^\alpha}\hat{q}_\alpha\right).
\end{equation}
The third Casimir operator is 
$\hat{K}_3 \sim \hat{K}_2\hat{B}+\cdots$.
We will consider a coset space $SU(2|1)/U(1)^2$
and this 
 defines a super sphere with four supercharges.
\par
Typical irreducible representations are characterized by
two quantum numbers, $(b,L)$ \cite{representation}.
The eigenvalues of the two Casimir operators are 
given by $K_2=L^2-b^2$ and $K_3=b(L^2-b^2)$.
(There are other types of irreducible representations but 
we do not consider them here.)
In terms of the $osp(1|2)$ subalgebra, this representation is
decomposed into two representations with superspin $L$
and $L-1/2.$ Hence the dimension of the irreducible representation
is $N=8L$. Any supermatrix with this size can be expanded in 
terms of polynomials generated by $\hat{l}_i$, $\hat{q}_{\alpha}$
and $\hat{\bar{q}}^{\alpha}$.  $\hat{B}$ can be solved by the 
third Casimir $\hat{K}_3$ and 
the polynomials do not depend on $\hat{B}$.  
Since three $\hat{l}_i$ satisfy
the constraint given by the second Casimir, this defines
two-dimensional supersphere with four grassmannian coordinates.
\par
In order to take a flat limit, we introduce the following superspace
coordinates,  
\begin{eqnarray}
 \begin{array}{ll}
  \displaystyle{\hat{x}_{i}=\left(\frac{\Theta}{L}\right)^{\frac{1}{2}}
   \hat{l}_{i}}, &
   \mbox{for } i=1, 2 \\
  \displaystyle{\hat{\theta}_1=\left(\frac{\Theta}{L}\right)^{\frac{1}{4}}
   \hat{q}_1,} &
   \displaystyle{\hat{\theta}_2=\left(\frac{\Theta}{L}\right)^{\frac{3}{4}}
   \hat{q}_2,} \\
  \displaystyle{\hat{\bar{\theta}^1}=
   \left(\frac{\Theta}{L}\right)^{\frac{3}{4}}\hat{\bar{q}^1}}, &
   \displaystyle{\hat{\bar{\theta}^2}=
   \left(\frac{\Theta}{L}\right)^{\frac{1}{4}}\hat{\bar{q}^2},} \\
  \displaystyle{\hat{b}=\left(\frac{1}{L}\right)\hat{B}.} &
 \end{array}
\end{eqnarray}
Again we need to take an asymmetric scaling for the fermionic
coordinates. 
The $\hat{l}_3$ is scaled as $L$ since we are considering
the vicinity of the north pole on the sphere.
In the large $L$ limit, the algebra among the coordinates becomes 
\begin{eqnarray}
 \begin{array}{ll}
  \displaystyle{\left[\hat{x}_+, \hat{x}_-\right]=2\Theta,} & \\
  \displaystyle{\left[\hat{x}_+, \hat{\theta}_1\right]=-\hat{\theta}_2,} & 
  \displaystyle{\left[\hat{x}_+, \hat{\bar{\theta}^2}\right]
  =\hat{\bar{\theta}^1},} \\
  \displaystyle{\left\{\hat{\theta}_1, \hat{\bar{\theta}^1}\right\}
   =(\hat{b}+1)\Theta,} &
   \displaystyle{\left\{\hat{\theta}_1, \hat{\bar{\theta}^2}\right\}
   =\hat{x}_-,} \\
  \displaystyle{\left\{\hat{\theta}_2, \hat{\bar{\theta}^2}\right\}
   =(\hat{b}-1)\Theta,} &
   \mbox{others} = 0,
 \end{array}
\end{eqnarray}
where $\hat{x}_\pm=\hat{x}_1 \pm i\hat{x}_2$.
Since $\hat{b}$ is a center of the algebra, we set it as a constant $b$.
Furthermore by the transformations 
\begin{eqnarray}
 \begin{array}{ll}
  \displaystyle{\hat{\varphi}_1=\hat{\theta}_1
   +\frac{1}{2\Theta}\hat{x}_-\hat{\theta}_2,} & 
   \displaystyle{\hat{\varphi}_2=\hat{\theta}_2,} \\
  \displaystyle{\hat{\bar{\varphi}^1}=\hat{\bar{\theta}^1},} &
   \displaystyle{\hat{\bar{\varphi}^2}=\hat{\bar{\theta}^2}
   -\frac{1}{2\Theta}\hat{x}_-\hat{\bar{\theta}^1},}
 \end{array}
\end{eqnarray}
this algebra reduces to a simpler noncommutative algebra
for the superspace coordinates;
\begin{eqnarray}
 && \left[\hat{x}_+, \hat{x}_-\right]=2\Theta, \nonumber \\
 && \left\{\hat{\varphi}_1, \hat{\bar{\varphi}^1}\right\}=(b+1)\Theta, 
  \nonumber \\
 && \left\{\hat{\varphi}_2, \hat{\bar{\varphi}^2}\right\}=(b-1)\Theta,
  \hspace{10mm}
  \mbox{others}=0.
\end{eqnarray}

Now we define the generators of supersymmetry in the flat limit as
\begin{eqnarray}
 && P_\pm=\pm \frac{1}{2}\mbox{adj}(\hat{x}_\pm), \nonumber \\
 && Q_\alpha=\mbox{adj}\left(\hat{\theta}_\alpha\right), \\
 && \bar{Q}^\alpha=\mbox{adj}\left(\hat{\bar{\theta}^\alpha}\right).
  \nonumber
\end{eqnarray} 
Then the following supersymmetry algebra holds
\begin{eqnarray}
 && \left[P_+, Q_1\right]=-\frac{1}{2}Q_2, \nonumber \\
 && \left[P_+, \bar{Q}^2\right]=\frac{1}{2}\bar{Q}^1, \\
 && \left\{Q_1, \bar{Q}^2\right\}=-2P_-, 
  \hspace{10mm}
  \mbox{others} = 0. \nonumber
\end{eqnarray}
There are four supercharges but
only half of them  generated by $Q_1$ and $\bar{Q}^2$
are dynamical supersymmetries. The other half are non-dynamical 
supersymmetries and do not generate space-time translation.
\par
We next introduce  additional operators  
\footnote{These operators are introduced so that they anticommute
with the supersymmetry generators in the flat limit. There might
be possible $1/L$ corrections to them before taking the flat limit.}
corresponding to
the generators eq.(\ref{addd}) in the $osp(1|2)$ case,
\begin{eqnarray}
 \hat{d}_\alpha &\equiv& \frac{1}{2L}
  \left\{{(\sigma_i)_\alpha}^\beta\left(\hat{l}_i\hat{q}_\beta 
				   + \hat{q}_\beta \hat{l}_i\right)
  -\left(\hat{B}\hat{q}_\alpha + \hat{q}_\alpha \hat{B}\right)\right\}, 
  \nonumber \\
 \hat{\bar{d}^\alpha} &\equiv& \frac{1}{2L}
  \left\{{(\sigma_i)_\beta}^\alpha\left(\hat{l}_i\hat{\bar{q}^\beta} 
				   + \hat{\bar{q}^\beta} \hat{l}_i\right)
  -\left(\hat{B}\hat{\bar{q}^\alpha} 
    + \hat{\bar{q}^\alpha} \hat{B}\right)\right\}.
\end{eqnarray} 
We can obtain covariant derivatives by taking the scaling limit of 
these generators, 
\begin{eqnarray}
 && \hspace*{-5mm}
  D_\alpha=\mbox{adj}\left(\hat{d}'_\alpha\right), \nonumber \\
 && \hspace*{-5mm}
  \bar{D}^\alpha=\mbox{adj}\left(\hat{\bar{d}'^\alpha}\right), 
  \nonumber \\
 && \hat{d}'_1=\left(\frac{\Theta}{L}\right)^{\frac{1}{4}}d_1
  =(1-b)\hat{\theta}_1 + \frac{1}{\Theta}x_-\hat{\theta}_2
  = (1-b)\hat{\varphi}_1 + \frac{1+b}{2\Theta}x_-\hat{\varphi}_2, 
  \nonumber \\
 && \hat{d}'_2=\left(\frac{\Theta}{L}\right)^{\frac{3}{4}}d_2
  =(1+b)\hat{\theta}_2
  =-(1+b)\hat{\varphi}_2, \\
 &&
  \hat{\bar{d}'^1}=\left(\frac{\Theta}{L}\right)^{\frac{3}{4}}\hat{\bar{d}^1} 
  =(1-b)\hat{\bar{\theta}^1}
  =(1-b)\hat{\bar{\varphi}^1},  \nonumber \\
 &&
  \hat{\bar{d}'^2}=\left(\frac{\Theta}{L}\right)^{\frac{1}{4}}\hat{\bar{d}^2}  
  =-(1+b)\hat{\bar{\theta}^2}+\frac{1}{\Theta}x_-\hat{\bar{\theta}^1}
  =-(1+b)\hat{\bar{\varphi}^2}+\frac{1-b}{2\Theta}x_-\hat{\bar{\varphi}^1}.
  \nonumber 
\end{eqnarray}
The covariant derivatives anticommute with $\hat{Q}_\alpha$ and 
$\hat{\bar{Q}^\alpha}$ and satisfy the following algebra,
\begin{eqnarray}
 && \left[P_+, D_1\right]=-\frac{1}{2}D_2, \nonumber \\
 && \left[P_+, \bar{D}^2\right]=\frac{1}{2}\bar{D}_1, \nonumber \\
 && \left\{D_1, \bar{D}^2\right\}=2(b^2-1)P_-. 
\end{eqnarray}

Functions on the superspace 
$(\hat{x}, \hat{\varphi}, \hat{\bar{\varphi}})$ 
are supermatrices and generally written as functions of supercoordinates:
$\Phi\left(\hat{x}_\pm, \hat{\varphi}_\alpha, 
\hat{\bar{\varphi}^\alpha}\right)$.
We can constrain the function by imposing the following constraints,
\begin{eqnarray}
 && D_2\Phi=-(1+b)\left[\hat{\varphi}_2, \Phi\right]=0, \nonumber \\ 
 && \bar{D}^1\Phi=(1-b)\left[\hat{\bar{\varphi}^1}, \Phi\right]=0.
\end{eqnarray}
These conditions automatically mean
\begin{eqnarray}
 Q_2\Phi=\bar{Q}^1\Phi=0,
\end{eqnarray}
and the superfield $\Phi$ depends only on 
$\left(\hat{x}_\pm, \hat{\varphi}_2, \hat{\bar{\varphi}^1}\right)$.
This is the chiral superfield 
\footnote{It would have been appropriate to parametrize the superspace
so that both of the two constraints are anti-chiral
and the constrained superfield becomes apparently chiral. 
It is merely a problem of notation.}
and supersymmetries are generated by
$Q_1$ and $\bar{Q}^2$ whose anticommutator becomes $P_-$.
As a result, on the chiral superfields, 
the algebra among generators for supersymmetries 
$Q_1$ and $\bar{Q}^2$, translations $P_\pm$, 
and covariant derivatives $D_1$ and $\bar{D}^2$ is given by 
\begin{eqnarray}
 && \left\{Q_1, \bar{Q}^2\right\}=-2P_-, \nonumber \\
 && \left\{D_1, \bar{D}^2\right\}=2(b^2-1)P_-, 
  \hspace{10mm} 
  \mbox{others}=0
\end{eqnarray}
They are written as differential operators;
\begin{eqnarray}
 && P_\pm = \Theta\frac{\partial}{\partial x_\mp}, \nonumber \\
 && Q_1 = (1+b)\Theta\frac{\partial}{\partial \bar{\varphi}^1}
  + \varphi_2\frac{\partial}{\partial x_+}, \nonumber \\
 && \bar{Q}^2 = -(1-b)\Theta\frac{\partial}{\partial \varphi_2}
  - \bar{\varphi}^1\frac{\partial}{\partial x_+}. \nonumber \\
 && D_1 = (1+b)\left[(1-b)\Theta\frac{\partial}{\partial \bar{\varphi}^1}
  -\varphi^2\frac{\partial}{\partial x_+} \right],
  \nonumber \\
 && \bar{D}^2 = (1-b)\left[(1+b)\Theta\frac{\partial}{\partial \varphi_2}
  - \bar{\varphi}^1\frac{\partial}{\partial x_+} \right]. \nonumber 
\end{eqnarray}
We can further constrain the superfield by $P_+$: 
\begin{equation}
 P_+ \Phi =0.
\end{equation}
Then the superfield becomes independent of the 
$\hat{x}_-$ coordinate and the system becomes essentially
one-dimensional system with two supersymmetries.

\section{Noncommutative superspace from psu(2$\mid$2)}
In this section we try to construct a noncommutative superspace 
in four dimensions based on $psu(2|2)$ graded algebra.
There are various possibilities for the scaling of grassmannian
coordinates when we take a flat limit.
We show two examples. 
More details will be discussed in a separate 
paper.

The $psu(2|2)$ graded algebra is given by
\begin{eqnarray}
 \begin{array}{ll}
  \displaystyle{
   \left[\hat{l}_i, \hat{l}_j\right]=i\epsilon_{ijk}\hat{l}_k,} & 
   \displaystyle{
   \left[\hat{\bar{l}}_i, \hat{\bar{l}}_j\right]
   =i\epsilon_{ijk}\hat{\bar{l}}_k,} \\
  \displaystyle{
   \left[\hat{l}_i, {\hat{q}_\alpha}^{\dot{\beta}}\right] 
   = -\frac{1}{2}{(\sigma_i)_\alpha}^\gamma
   {\hat{q}_\gamma}^{\dot{\beta}},} &
   \displaystyle{
   \left[\hat{l}_i, {\hat{\bar{q}}}_{\dot{\alpha}}^{\beta}\right]
   = \frac{1}{2}{(\sigma_i)_\gamma}^\beta 
   {\hat{\bar{q}}_{\dot{\alpha}}}^{\gamma},} \\
  \displaystyle{
   \left[\hat{\bar{l}}_i, {\hat{q}_\alpha}^{\dot{\beta}}\right] 
   = \frac{1}{2}{(\sigma_i)_{\dot{\gamma}}}^{\dot{\beta}}
   {\hat{q}_\alpha}^{\dot{\gamma}},} &
   \displaystyle{
   \left[\hat{\bar{l}}_i, {\hat{\bar{q}}}_{\dot{\alpha}}^{\beta}\right]
   = -\frac{1}{2}{(\sigma_i)_{\dot{\alpha}}}^{\dot{\gamma}} 
   {\hat{\bar{q}}_{\dot{\gamma}}}^{\beta},} \\
  \displaystyle{
   \left\{{\hat{q}_\alpha}^{\dot{\beta}}, 
    {\hat{\bar{q}}_{\dot{\gamma}}}^\delta\right\}
   = \delta_{\dot{\gamma}}^{\dot{\beta}}{(\sigma_i)_\alpha}^\delta
   \hat{l}_i -
   \delta_\alpha^\delta{(\sigma_i)_{\dot{\gamma}}}^{\dot{\beta}}
   \hat{\bar{l}}_i.} &
 \end{array}
\end{eqnarray}
The bosonic part of $su(2|2)$ consists of two sets of
$su(2)$ algebra generated by
$\hat{l}_i$ and $\hat{\bar{l}}_i$ and eight odd generators transform 
as spinors both under two $su(2)$'s.
Based on this graded algebra we can construct a four-dimensional 
superspace $PSU(2|2)/U(1)^2$.

\subsection{$d=4$ flat noncommutative superspace}
In order to take a flat limit we first consider the following scaling of
the superspace coordinates, 
\begin{eqnarray} 
 && \hat{x}_i=\left(\frac{\Theta}{L}\right)^{\frac{1}{2}}\hat{l}_i, 
  \hspace{10mm}
  \hat{\bar{x}}_i
  =\left(\frac{\Theta}{L}\right)^{\frac{1}{2}}\hat{\bar{l}}_i, 
  \hspace{10mm}
  \mbox{for } i=1, 2 \nonumber \\
 && \left(\hat{\theta}_2^{\dot{1}}, \ \hat{\bar{\theta}}_{\dot{1}}^1,\  
     \hat{\bar{\theta}}_{\dot{2}}^1, \ \hat{\bar{\theta}}_{\dot{2}}^2\right)
 =\left(\frac{\Theta}{L}\right)^{\frac{1}{4}}
 \left(\hat{q}_2^{\dot{1}}, \ \hat{\bar{q}}_{\dot{1}}^1,\  
  \hat{\bar{q}}_{\dot{2}}^1, \ \hat{\bar{q}}_{\dot{2}}^2\right),
 \nonumber \\
&& \left(\hat{\theta}_1^{\dot{1}}, \ \hat{\theta}_1^{\dot{2}},\  
     \hat{\theta}_2^{\dot{2}}, \ \hat{\bar{\theta}}_{\dot{1}}^2\right)
 =\left(\frac{\Theta}{L}\right)^{\frac{3}{4}}
 \left(\hat{q}_1^{\dot{1}}, \ \hat{q}_1^{\dot{2}},\  
  \hat{q}_2^{\dot{2}}, \ \hat{\bar{q}}_{\dot{1}}^2\right).
\end{eqnarray}
We use a similar asymmetric scaling of the fermionic coordinates to 
the two-dimensional cases.
The $\hat{l}_3$ and $\hat{\bar{l}_3}$ are scaled as $L$ and $cL$ 
respectively where $c$ is an arbitrary constant. 
Taking the large $L$ limit the algebra among the coordinates becomes
\begin{eqnarray}
 \begin{array}{ll}
  \left[\hat{x}_+, \hat{x}_-\right]=2\Theta, & 
   \left[\hat{\bar{x}}_+, \hat{\bar{x}}_-\right]=2c\Theta, \\
  \left[\hat{x}_-, \hat{\theta}_2^{\dot{1}}\right]
   =-\hat{\theta}_1^{\dot{1}}, &
   \left[\hat{x}_-, \hat{\bar{\theta}}_{\dot{1}}^1\right]
   =\hat{\bar{\theta}}_{\dot{1}}^2 \\
  \left[\hat{\bar{x}}_-, \hat{\theta}_2^{\dot{1}}\right]
   =\hat{\theta}_2^{\dot{2}}, &
   \left[\hat{\bar{x}}_-, \hat{\bar{\theta}}_{\dot{2}}^2\right]
   =-\hat{\bar{\theta}}_{\dot{1}}^2, \\
  \left\{\hat{\theta}_2^{\dot{1}}, \hat{\bar{\theta}}_{\dot{1}}^1\right\}
   =\hat{x}_+, &
   \left\{\hat{\theta}_2^{\dot{1}},
    \hat{\bar{\theta}}_{\dot{2}}^2\right\}=-\hat{\bar{x}}_+, \\
  \left\{\hat{\theta}_1^{\dot{1}},
   \hat{\bar{\theta}}_{\dot{1}}^1\right\}=(1-c)\Theta, &
   \left\{\hat{\theta}_1^{\dot{2}},
    \hat{\bar{\theta}}_{\dot{2}}^1\right\}=(1+c)\Theta, \\
  \left\{\hat{\theta}_2^{\dot{1}},
   \hat{\bar{\theta}}_{\dot{1}}^2\right\}=-(1+c)\Theta, &
  \left\{\hat{\theta}_2^{\dot{2}},
   \hat{\bar{\theta}}_{\dot{2}}^2\right\}=-(1-c)\Theta, \\
  \mbox{others}=0. &
 \end{array}
\end{eqnarray}
By introducing the following fermionic coordinates 
\begin{eqnarray}
 && \hat{\varphi}_2^{\dot{1}}=\hat{\theta}_2^{\dot{1}}
  -\frac{1}{2\Theta}\hat{x}_+\hat{\theta}_1^{\dot{1}}
  +\frac{1}{2c\Theta}\hat{\bar{x}}_+\hat{\theta}_2^{\dot{2}}, 
  \nonumber \\ 
 && \hat{\bar{\varphi}}_{\dot{1}}^1=\hat{\bar{\theta}}_{\dot{1}}^1
  +\frac{1}{2\Theta}\hat{x}_+\hat{\bar{\theta}}_{\dot{1}}^2, \\
 && \hat{\bar{\varphi}}_{\dot{2}}^2=\hat{\bar{\theta}}_{\dot{2}}^2
  -\frac{1}{2c\Theta}\hat{\bar{x}}_+\hat{\bar{\theta}}_{\dot{1}}^2, 
  \nonumber \\
 && \hat{\varphi}_\alpha^{\dot{\beta}}=\hat{\theta}_\alpha^{\dot{\beta}}, 
  \qquad
  \hat{\bar{\varphi}}_{\dot{\alpha}}^\beta
  =\hat{\bar{\theta}}_{\dot{\alpha}}^\beta, 
  \hspace{10mm} \mbox{for others} \nonumber 
\end{eqnarray}
this algebra is much simplified to satisfy the canonical forms;
\begin{eqnarray}
 \begin{array}{ll}
  \left[\hat{x}_+, \hat{x}_-\right]=2\Theta, & 
   \left[\hat{\bar{x}}_+, \hat{\bar{x}}_-\right]=2c\Theta, \\
  \left\{\hat{\varphi}_1^{\dot{1}},
   \hat{\bar{\varphi}}_{\dot{1}}^1\right\}
  =(1-c)\Theta, & 
  \left\{\hat{\varphi}_1^{\dot{2}},
   \hat{\bar{\varphi}}_{\dot{2}}^1\right\}
  =(1+c)\Theta, \\
  \left\{\hat{\varphi}_2^{\dot{1}},
   \hat{\bar{\varphi}}_{\dot{1}}^2\right\}
  =-(1+c)\Theta, & 
  \left\{\hat{\varphi}_2^{\dot{2}},
   \hat{\bar{\varphi}}_{\dot{2}}^2\right\}
  =-(1-c)\Theta. \\
 \end{array}
\end{eqnarray}
We define the generators of supersymmetry in the flat limit as
\begin{eqnarray}
 && P_\pm = \pm \frac{1}{2}\mbox{adj} \ x_\pm, 
  \hspace{10mm}
 \bar{P}_\pm = \pm \frac{1}{2}\mbox{adj} \ \bar{x}_\pm, \nonumber \\
 && Q_{\alpha}^{\dot{\beta}} = \mbox{adj} \ \theta_{\alpha}^{\dot{\beta}}, 
  \hspace{17mm}
  \bar{Q}_{\dot{\alpha}}^\beta=
  \mbox{adj} \ \bar{\theta}_{\dot{\alpha}}^\beta. 
\end{eqnarray}
Then the supersymmetry algebra has the following forms,
\begin{eqnarray}
 \label{su22-PQ}
 \begin{array}{ll}
  \displaystyle{\left[P_-, Q_2^{\dot{1}}\right]=\frac{1}{2}Q_1^{\dot{1}},} &
   \displaystyle{\left[P_-, \bar{Q}_{\dot{1}}^1\right]
   =-\frac{1}{2}\bar{Q}_{\dot{1}}^2,}  \\
  \displaystyle{\left[\bar{P}_-, Q_2^{\dot{1}}\right]
   =-\frac{1}{2}Q_2^{\dot{2}},} &
   \displaystyle{\left[\bar{P}_-, \bar{Q}_{\dot{2}}^2\right]
   =\frac{1}{2}\bar{Q}_{\dot{1}}^2,} \\
  \displaystyle{\left\{Q_2^{\dot{1}}, \bar{Q}_{\dot{1}}^1\right\}
   =2P_+,} &
   \left\{Q_2^{\dot{1}}, \bar{Q}_{\dot{2}}^2\right\}=-2\bar{P}_+, \\
   \mbox{others}=0. &
\end{array}
\end{eqnarray}
Only three supercharges $Q_2^{\dot{1}}, \bar{Q}_{\dot{1}}^1$ and 
$\bar{Q}_{\dot{2}}^2$ generate dynamical supersymmetries, i.e., 
anticommutators among them become the generators of the space-time 
translations. The other supercharges are generators of non-dynamical
supersymmetries. 
In order to construct covariant derivatives, it is useful to consider 
the following operators, 
\begin{eqnarray}
 && \hat{d}_\alpha^{\dot{\beta}}=-\frac{1}{2L}
  \left\{{(\sigma_i)_\alpha}^\gamma
   \left(\hat{l}_i\hat{q}_\gamma^{\dot{\beta}}
    +\hat{q}_\gamma^{\dot{\beta}}\hat{l}_i\right)
  +{(\sigma_i)_{\dot{\gamma}}}^{\dot{\beta}}
  \left(\hat{\bar{l}}_i\hat{q}_\alpha^{\dot{\gamma}}
   +\hat{q}_\alpha^{\dot{\gamma}}\hat{\bar{l}}_i\right)\right\} \\
 && \hat{\bar{d}}_{\dot{\alpha}}^\beta=\frac{1}{2L}
  \left\{{(\sigma_i)_\gamma}^\beta
   \left(\hat{l}_i\hat{\bar{q}}_{\dot{\alpha}}^\gamma
    +\hat{\bar{q}}_{\dot{\alpha}}^\gamma\hat{l}_i\right)
  +{(\sigma_i)_{\dot{\alpha}}}^{\dot{\gamma}}
  \left(\hat{\bar{l}}_i\hat{\bar{q}}_{\dot{\gamma}}^\beta
   +\hat{\bar{q}}_{\dot{\gamma}}^\beta\hat{\bar{l}}_i\right)\right\}. 
\end{eqnarray}
We can obtain the covariant derivatives which anticommute with 
the super charges in the flat limit,
\begin{eqnarray}
 &&
 D_1^{\dot{1}}=\left(\frac{\Theta}{L}\right)^{\frac{3}{4}}
 \mbox{adj} \ d_1^{\dot{1}}
 =-(1+c)\mbox{adj} \ \hat{\theta}_1^{\dot{1}}, \nonumber \\
 && D_1^{\dot{2}}=\left(\frac{\Theta}{L}\right)^{\frac{3}{4}}
  \mbox{adj}\ d_1^{\dot{2}}
  =-(1-c)\mbox{adj} \ \hat{\theta}_1^{\dot{2}}, \nonumber \\
 && D_2^{\dot{1}}=\left(\frac{\Theta}{L}\right)^{\frac{1}{4}}
  \mbox{adj}\ d_2^{\dot{1}}
  =\mbox{adj} \left[ (1-c)\hat{\theta}_2^{\dot{1}}
  -\frac{1}{\Theta}\hat{x}_+\theta_1^{\dot{1}}
  -\frac{1}{\Theta}\hat{\bar{x}}_+\hat{\theta}_2^{\dot{2}}\right] 
  \nonumber \\
 && \hspace{6.7mm} =\mbox{adj} \left[ (1-c)\hat{\varphi}_2^{\dot{1}}
   -\frac{1+c}{2\Theta}\hat{x}_+\hat{\varphi}_1^{\dot{1}}
  -\frac{1+c}{2\Theta c}\hat{\bar{x}}_+
  \hat{\bar{\varphi}}_2^{\dot{2}}\right], \nonumber \\
  && D_2^{\dot{2}}=\left(\frac{\Theta}{L}\right)^{\frac{3}{4}}
 \mbox{adj} \ d_2^{\dot{2}}
 =(1+c)\mbox{adj} \ \hat{\theta}_2^{\dot{2}}, \nonumber \\
 && \bar{D}_{\dot{1}}^1=\left(\frac{\Theta}{L}\right)^{\frac{1}{4}}
  \mbox{adj} \ \hat{\bar{d}}_{\dot{1}}^1 
  =\mbox{adj} \left[(1+c)\hat{\bar{\theta}}_{\dot{1}}^1
	      +\frac{1}{\Theta}\hat{x}_+\hat{\bar{\theta}}_{\dot{1}}^2\right] 
  \\
  && \hspace{6.7mm}=\mbox{adj}
  \left[(1+c)\hat{\bar{\varphi}}_{\dot{1}}^1
  +\frac{1-c}{2\Theta}\hat{x}_+
  \hat{\bar{\varphi}}_{\dot{1}}^2\right], \nonumber \\
 && \bar{D}_{\dot{1}}^2
  =\left(\frac{\Theta}{L}\right)^{\frac{3}{4}}
  \mbox{adj} \ \hat{\bar{d}}_{\dot{1}}^2
  =-(1-c)\mbox{adj} \ \hat{\bar{\theta}}_{\dot{1}}^2
  =-(1-c)\mbox{adj} \ \hat{\bar{\varphi}}_{\dot{1}}^2,
 \nonumber \\
 && \bar{D}_{\dot{2}}^1 
  =\left(\frac{\Theta}{L}\right)^{\frac{1}{4}}
  \mbox{adj} \ \hat{\bar{d}}_{\dot{2}}^1
  =(1-c)\mbox{adj} \ \hat{\bar{\theta}}_{\dot{2}}^1
  =(1-c)\mbox{adj} \ \hat{\bar{\varphi}}_{\dot{2}}^1,
  \nonumber \\
 && \bar{D}_{\dot{2}}^2
  =\left(\frac{\Theta}{L}\right)^{\frac{1}{4}}
  \mbox{adj} \ \hat{\bar{d}}_{\dot{2}}^2
  =\mbox{adj} \ \left[-(1+c)\hat{\bar{\theta}}_{\dot{2}}^2
	+\frac{1}{\Theta}\hat{\bar{x}}_+\hat{\bar{\theta}}_{\dot{1}}^2\right]
  \nonumber \\
 && \hspace{6.7mm}=\mbox{adj}
  \left[-(1+c)\hat{\bar{\varphi}}_{\dot{2}}^2
  -\frac{1-c}{2\Theta c}\hat{\bar{x}}_+
  \hat{\bar{\varphi}}_{\dot{1}}^2\right]. \nonumber 
\end{eqnarray}
The covariant derivatives and the generators of space-time translations 
satisfy the following algebra,
\begin{eqnarray}
 \label{su22-PD}
  \begin{array}{ll}
   \displaystyle{
    \left[P_-, D_2^{\dot{1}}\right]=\frac{1}{2}D_1^{\dot{1}},} &
    \displaystyle{\left[P_-, \bar{D}_{\dot{1}}^1\right]
    =-\frac{1}{2}\bar{D}_{\dot{1}}^2,} \\
   \displaystyle{\left[\bar{P}_-, D_2^{\dot{1}}\right]
    =-\frac{1}{2}D_2^{\dot{2}},} &
    \displaystyle{\left[\bar{P}_-, \bar{D}_{\dot{2}}^2\right]
    =\frac{1}{2}\bar{D}_{\dot{1}}^2,} \\
   \displaystyle{\left\{D_2^{\dot{1}}, \bar{D}_{\dot{1}}^1\right\}
    =-2(1-c^2)P_+,} &
    \displaystyle{\left\{D_2^{\dot{1}}, \bar{D}_{\dot{2}}^2\right\}
    =-2(1-c^2)\bar{P}_+,} \\
   \mbox{others}=0.&
  \end{array}
\end{eqnarray}
Field theories based on this algebra in the flat limit 
have two-dimensional like supersymmetries 
because only $P_+$ and $\bar{P}_+$ appear 
in the right hand sides of the algebra (\ref{su22-PQ}) and
(\ref{su22-PD}).  

Next we consider another choice of scaling for the fermionic superspace 
coordinates,
\begin{eqnarray}
 && \left(\hat{\theta}_2^{\dot{1}}, \ \hat{\theta}_1^{\dot{2}},\  
     \hat{\bar{\theta}}_{\dot{1}}^1, \ \hat{\bar{\theta}}_{\dot{2}}^2\right)
 =\left(\frac{\Theta}{L}\right)^{\frac{1}{4}}
 \left(\hat{q}_2^{\dot{1}}, \ \hat{q}_1^{\dot{2}},\  
  \hat{\bar{q}}_{\dot{1}}^1, \ \hat{\bar{q}}_{\dot{2}}^2\right),
 \nonumber \\
&& \left(\hat{\theta}_1^{\dot{1}}, \ \hat{\theta}_2^{\dot{2}},\  
     \hat{\bar{\theta}}_{\dot{2}}^1, \ \hat{\bar{\theta}}_{\dot{1}}^2\right)
 =\left(\frac{\Theta}{L}\right)^{\frac{3}{4}}
 \left(\hat{q}_1^{\dot{1}}, \ \hat{q}_2^{\dot{2}},\  
  \hat{\bar{q}}_{\dot{2}}^1, \ \hat{\bar{q}}_{\dot{1}}^2\right). 
\end{eqnarray}
Then in the large $L$ limit the coordinates of the superspace satisfy 
\begin{eqnarray}
\begin{array}{ll}
 \left[\hat{x}_+, \hat{x}_-\right]=2\Theta,& 
  \left[\hat{\bar{x}}_+, \hat{\bar{x}}_-\right]=2c\Theta,\\
 \left[\hat{x}_+,
  \hat{\theta}_1^{\dot{2}}\right]=-\hat{\theta}_2^{\dot{2}}, &
 \left[\hat{x}_-,
  \hat{\theta}_2^{\dot{1}}\right]=-\hat{\theta}_1^{\dot{1}}, \\
 \left[\hat{x}_+,
  \hat{\bar{\theta}}_{\dot{2}}^2\right]=\hat{\bar{\theta}}_{\dot{2}}^1,&
 \left[\hat{x}_-,
  \hat{\bar{\theta}}_{\dot{1}}^1\right]=\hat{\bar{\theta}}_{\dot{1}}^2,\\
 \left[\hat{\bar{x}}_+,
  \hat{\theta}_1^{\dot{2}}\right]=\hat{\theta}_1^{\dot{1}}, &
 \left[\hat{\bar{x}}_-,
  \hat{\theta}_2^{\dot{1}}\right]=\hat{\theta}_2^{\dot{2}}, \\
 \left[\hat{\bar{x}}_+,
  \hat{\bar{\theta}}_{\dot{1}}^1\right]=-\hat{\bar{\theta}}_{\dot{2}}^1,&
 \left[\hat{\bar{x}}_-,
  \hat{\bar{\theta}}_{\dot{2}}^2\right]=-\hat{\bar{\theta}}_{\dot{1}}^2,\\
 \left\{\hat{\theta}_1^{\dot{2}}, \hat{\bar{\theta}}_{\dot{1}}^1\right\}
  =-\hat{\bar{x}}_-, &
  \left\{\hat{\theta}_1^{\dot{2}}, \hat{\bar{\theta}}_{\dot{2}}^2\right\}
  =\hat{x}_-, \\
 \left\{\hat{\theta}_2^{\dot{1}}, \hat{\bar{\theta}}_{\dot{1}}^1\right\}
  =\hat{x}_+, &
  \left\{\hat{\theta}_2^{\dot{1}}, \hat{\bar{\theta}}_{\dot{2}}^2\right\}
  =-\hat{\bar{x}}_+, \\
 \left\{\hat{\theta}_1^{\dot{1}}, \hat{\bar{\theta}}_{\dot{1}}^1\right\}
  =\Theta (1-c),&
  \left\{\hat{\theta}_2^{\dot{1}}, \hat{\bar{\theta}}_{\dot{1}}^2\right\}
  =-\Theta (1+c), \\
 \left\{\hat{\theta}_1^{\dot{2}}, \hat{\bar{\theta}}_{\dot{2}}^1\right\}
  =\Theta (1+c), &
  \left\{\hat{\theta}_2^{\dot{2}}, \hat{\bar{\theta}}_{\dot{2}}^2\right\}
  =-\Theta (1-c).
\end{array} 
\end{eqnarray} 
We define the generators of supersymmetries and space-time translations as 
$Q_\alpha^{\dot{\beta}}=\mbox{adj} \ \hat{\theta}_\alpha^{\dot{\beta}}$, 
$\bar{Q}_{\dot{\alpha}}^\beta=\mbox{adj} \ \hat{\bar{\theta}}_{\dot{\alpha}}^\beta$, 
$P_\pm=\pm\frac{1}{2}\mbox{adj} \ \hat{x}_\pm$ and 
$\bar{P}_\pm=\pm\frac{1}{2}\mbox{adj} \ \hat{\bar{x}}_\pm$.
In this case, it can be easily seen  that the anticommutators among 
four supercharges 
$Q_1^{\dot{2}}, ~Q_2^{\dot{1}}, ~\bar{Q}_{\dot{1}}^1$ and $ \bar{Q}_{\dot{2}}^2$ 
become four space-time translations $P_\pm$ and $\bar{P}_\pm$.
Therefore noncommutative theories based on this algebra have 
four dynamical supersymmetries.

\subsection{Seiberg's noncommutative superspace as a constrained system}
Here we briefly explain that the Seiberg's noncommutative superspace
\cite{Seiberg} can be understood as a constrained system and the
noncommutative algebra for the coordinates is realized by
the Dirac bracket under the constraints.
\par
We begin with a little more general setting.
We consider supercovariant derivatives (\ref{1.1}, \ref{1.2})
with the following gauge field backgrounds,
\begin{eqnarray}
 && D_\mu=\partial_\mu-\frac{i}{2}f_{\mu\nu}x^\nu, \nonumber \\
 && D_\alpha=\frac{\partial}{\partial \theta^\alpha}
  -i(\sigma^\mu \bar{\theta})_\alpha \partial_\mu
  +\frac{i}{2}f_{\alpha\beta}\theta^\beta, \\
 && \bar{D}_{\dot{\alpha}}=
  -\frac{\partial}{\partial \bar{\theta}^{\dot{\alpha}}}
  +i(\theta\sigma^\mu)_{\dot{\alpha}} \partial_\mu
  -\frac{i}{2}f_{\dot{\alpha}\dot{\beta}}\theta^{\dot{\beta}}.
\end{eqnarray}
They are natural generalization of the bosonic covariant
derivatives in a constant magnetic field, but the field 
strengths depend on the superspace coordinates;
\begin{eqnarray}
  \begin{array}{lll}
   \displaystyle{F_{\mu\nu}=f_{\mu\nu}} & 
    \displaystyle{F_{\mu\alpha}=
    -\frac{i}{2}f_{\mu\nu}(\sigma^\nu\bar{\theta})_\alpha,} & 
    \displaystyle{F_{\mu\dot{\alpha}}=
    \frac{i}{2}f_{\mu\nu}(\theta\sigma^\nu)_{\dot{\alpha}}},\\
   \displaystyle{F_{\alpha\beta}=f_{\alpha\beta},} &
    \displaystyle{F_{\alpha\dot{\beta}}
    =if_{\mu\nu}(\sigma^\mu)_{\alpha\dot{\beta}}x^\nu,} & 
    \displaystyle{F_{\dot{\alpha}\dot{\beta}}
    =f_{\dot{\alpha}\dot{\beta}}}.
  \end{array}
\end{eqnarray}
Imposing $D_\alpha=\bar{D}_{\dot{\alpha}}=D_\mu =0$ 
as the second class constraints, we can calculate
the Dirac brackets of the superspace coordinates and obtain
the noncommutative algebra similar to that given in \cite{Klemm}.

In the following we consider an easier case for simplicity, 
namely, a case where only $f_{\alpha\beta}$ are nonvanishing. 
We then impose the second class constrain by only $D_{\alpha}$'s
and  calculate the Dirac bracket 
\begin{equation}
  [A, B\}_D=[A, B\}+ i[A, D_\alpha\}f^{-1}_{\alpha\beta}[D_\beta, B\} 
\end{equation}
as
\begin{eqnarray}
 && \left[x^\mu, x^\nu\right]_D
  =i f^{-1}_{\alpha\beta}\left(\sigma^\mu\bar{\theta}\right)_\alpha
  \left(\sigma^\nu\bar{\theta}\right)_\beta, \nonumber \\
  && \left[x^\mu, \theta_\alpha\right]_D=
   -f^{-1}_{\alpha\beta}\left(\sigma^\mu\bar{\theta}\right)_\beta, \\
  && \left\{\theta_\alpha, \theta_\beta\right\}_D
   =i f^{-1}_{\alpha\beta}. \nonumber
\end{eqnarray}
This is nothing but the noncommutative algebra given in 
the paper by Seiberg \cite{Seiberg}. 
The noncommutative parameter $C_{\alpha \beta}$ in \cite{Seiberg}
is related to our $f$ as
$C_{\alpha \beta}= i f_{\alpha \beta}^{-1}$.
If we redefine the coordinate as
\begin{equation}
 y^{\mu}= x^{\mu} + i \theta \sigma^{\mu} \bar{\theta},
\label{redefy}
\end{equation}
they become commutable with the other coordinates
\begin{equation}
[y^{\mu}, y^{\nu}]_D =[y^{\mu}, \theta^{\alpha}]_D = 
[y^{\mu}, \bar{\theta}^{\dot{\alpha}}]_D=0 
\end{equation}

In this case where only $f_{\alpha \beta}$ is nonvanishing,
we can more easily obtain the 
canonical pairs on the reduced superspace
 by a similarity transformation 
\begin{equation}
 {\cal O} \rightarrow \exp(-i \theta \sigma^{\mu} \bar{\theta} \partial_{\mu})
  \ {\cal O} \ \exp(i \theta \sigma^{\mu} \bar{\theta} \partial_{\mu}).
\end{equation}
Then the mutually commutative set of supercovariant derivatives $D_{\alpha}$ 
and global charges $q_{\alpha}$ are given by
\begin{eqnarray}
\label{confusing}
 D_{\alpha} &=& \frac{\partial}{\partial \theta^\alpha}
  +\frac{i}{2}f_{\alpha\beta}\theta^\beta, \\
 q_{\alpha} &=& i\frac{\partial}{\partial\theta^\alpha}
  -2 (\sigma^\mu\bar{\theta})_\alpha\partial_\mu
  +\frac{1}{2}f_{\alpha\beta}\theta^\beta.
\end{eqnarray}
The remaining coordinates on the phase space which (anti)commute
with $D_{\alpha}$ are
\begin{equation}
\psi_{\alpha} = f_{\alpha \beta}^{-1} 
 (q_{\beta} + 2 (\sigma^\mu\bar{\theta})_\beta \partial_\mu)
 = i f_{\alpha \beta}^{-1} D_{\beta}+ \theta_{\alpha} 
\end{equation}
and $(x^{\mu}, \partial/\partial x^{\mu}, \bar{\theta}^{\dot{\alpha}},
\partial/\partial \bar{\theta}^{\dot{\alpha}})$.
They satisfy the canonical algebra with $\psi$
\begin{equation}
\{ \psi_{\alpha}, \psi_{\beta}\} = i f_{\alpha \beta}^{-1}.
\end{equation}
\par
The set $(\psi_{\alpha},
x^{\mu}, \partial/\partial x^{\mu}, \bar{\theta}^{\dot{\alpha}},
\partial/\partial \bar{\theta}^{\dot{\alpha}})$
gives the phase space coordinates for the constrained system we are
considering now.
When we construct a field theory on the noncommutative space
$(x^{\mu},\psi_{\alpha},\bar{\theta}^{\dot{\alpha}})$,
we can introduce the canonical conjugate to $\psi$
as the adjoint action 
$$
\partial/\partial \psi_{\alpha}=-i \mbox{ adj } q_{\alpha}.
$$
Then the covariant derivatives\footnote{
It is confusing to use the same word as the operator defined in
(\ref{confusing}). The covariant derivative in (\ref{confusing})
is an operator used to define the noncommutative superspace 
as a constrained system. The covariant derivative here is
an operator acting within the constrained space that anticommutes
with the supercharges.
}
and the supercharges can be
defined in the same way as that given in \cite{Seiberg}.


\indent
\section{Conclusions and Discussions}
In this paper, we have constructed noncommutative superspaces 
based on graded (super) Lie algebras. In particular, we consider
fuzzy supersphere based on $osp(1|2)$ and $su(2|1)$ algebras.
They give two-dimensional supersphere with two and four real
supercharges. We then consider flat limits. In order to take flat
limits with the fermionic noncommutativity, we needed to take an 
asymmetric scaling limit for fermionic coordinates on superspace.
We also obtained covariant derivatives and imposed chiral constraints
to remove half degrees of freedom.  
This method was  generalized to four dimensional
noncommutative superspaces based on $psu(2|2)$ algebra \cite{HIU22}. 
In this case, there are varieties to assign scalings to the 
fermionic coordinates when we take a flat limit.
We showed two examples. One is similar to the two-dimensional
cases and we have obtained supersymmetry generators and 
covariant derivatives. 
This system is two-dimensional like in a sense that only two
generators of space-time translation appear in the anticommutators
of the supersymmetry generators.
The other example is more nontrivial.
With this scaling, space-time translation generators 
 into the all four directions appear.
More details are left for future investigations.
It would be also interesting to investigate
other scaling limits of the $psu(2|2)$ or $su(2|2)$ algebras
which can give the noncommutative superspace  given in the paper
\cite{BS}.

\par
We have also investigated these noncommutative superspaces 
as constrained systems. This is an analogue of the lowest Landau
level system of particles moving in a constant magnetic field.
We obtained two sets of operators, super covariant derivatives and 
super guiding center coordinates. 
They are obtained by the right and 
the left multiplications on the group manifolds. So they are
commutative to each other. The lowest Landau level conditions 
are generalized by adding fermionic constrains in addition to 
the ordinary bosonic condition for the lowest Landau level states.
Imposing  gauge fixing conditions, 
we calculated the Dirac brackets for superspace coordinates to
obtain noncommutative superspace coordinates.
This method can be extended to more general cases to obtain 
more general noncommutative superspaces. 
We want to report it in a future publication. 
Along this line, it will also be interesting to investigate a
supersymmetric generalization of $W_{\infty}$ algebraic 
structure, which plays an important role to study the physics of 
the lowest Landau level systems \cite{IKS}. 
\par
Our construction has an advantage that Jacobi identities
of the supersymmetries and the associativities of the star 
products are manifest, but it is restricted to the noncommutative
superspaces with Lie algebraic noncommutativity.
Namely the supersymmetry algebras satisfy Lie algebras.
In the papers \cite{OV,Seiberg}, the anticommutators for
supersymmetries contain the second order derivative operators.
In order to construct these structures based on a supermatrix approach,
we may need to investigate supermatrix models without the graded 
Lie algebraic structures.

{\bf Acknowledgments}\par
We would like to thank Drs. H. Fuji, T. Kimura, Y. Kitazawa,
M. Sakaguchi and T. Suyama for discussions.
The work of H.U. is supported in part by JSPS Research Fellowships for
Young Scientists. The work of S. I. is supported in part by
the Grant-in-Aid for Scientific Research from the Ministry of
Education, Science and Culture of Japan.
\appendix
\section{Cartan one-forms and the generalized In$\ddot{\rm{\bf o}}$n$\ddot{\rm{\bf u}}$-Wigner contraction}
In this appendix, we briefly explain the method of Cartan one-forms
to obtain the left and right multiplications. Then we explain 
the generalized In$\ddot{\rm o}$n$\ddot{\rm u}$-Wigner contraction proposed in \cite{gIW}.
\par
Suppose we have a Lie algebra  $[T_a, T_b]=f_{ab}{}^c T_c$.
A group manifold generated by this algebra can be parametrized
as $g=e^{ix^aT_a}$. 
Cartan one-forms $e^{a}$ are defined by $g^{-1} dg =ie^aT_a $
and satisfy the Maure-Cartan (MC) equation $de^a = f_{bc}^{a} e^b e^c/(2i)$.
If we write them 
as $e^a=d x^m e_m{}^a(x)$, 
covariant derivatives $D_a= (e^{-1})_a^m \partial_m/i$ 
generate the right multiplication $g \rightarrow gh$ and obey
the Lie algebra $[D_a, D_b]= f_{ab}{}^c D_c$. 
The left multiplication generators are 
similarly obtained from $dg g^{-1}$
and commute with the right multiplications.
\par

The generalized IW contraction can be obtained as follows.
We rescale the parameter $x^a$ on the group manifold as
$x^a \rightarrow s^{n_a} x^a$ and take $s \rightarrow 0$ limit.
Since the Cartan one-form $e^a$ is written in terms of a
polynomial of $x$'s, it can be expanded by $s$ as
\begin{equation}
e_a=\sum_{n} s^n e_{a[n]}~~~. 
\end{equation}
Here we interpret that each $e_{a[n]}$ is a different
Cartan one-form corresponding to different generators $T_{a[n]}$.
In this sense, this is an expansion \cite{az} rather than 
a contraction \cite{IW,HKSpp}.
MC equations are 
satisfied order by order
\begin{equation}
de_{a[n]} = \frac{1}{2i} {f^{bc}}_{a} e_{b[l]} e_{c[n-l]}
\label{MC}
\end{equation}
and they determine commutation relations between the
generators $T_{a[n]}$.
It is obvious from the MC equations (\ref{MC})
that the commutation relations are closed among generators
with weights $[n]$ less than some fixed number.
Jacobi identities are automatically satisfied.

\section{Supercovariant derivatives }
In this appendix, we give a derivation of (\ref{covosp}).
An group element of the noncommutative super-translation group
generated by $T_A=\{L_x,L_y,L_z,Q_\pm\}$
satisfying \bref{covosp} 
is parameterized as $g(x,y,\phi,\theta^+,\theta^-)$.
Left invariant Cartan 1-forms are obtained by
$g^{-1}dg=dz^ME_M{}^AT_A$  as
\bea
{\bf E}^x&=&dx-\frac{i}{2}d\theta^+\theta^+\nn\\
{\bf E}^y&=&dy+\frac{1}{2}d\theta^+\theta^+\nn\\
{\bf E}^z&=&\left(d\phi-\frac{1}{2}(dx~y-dy~x)\right)
+\frac{i}{2}(d\theta^+\theta^-+d\theta^-\theta^+)\nn\\
E^+&=&d\theta^+\nn\\
E^-&=&d\theta^-+\frac{i}{2}(dx+idy)\theta^+\label{C1form}~~~.
\eea
The coefficients of the Cartan 1-forms are give as
\bea
E_M{}^A=\left(
\begin{array}{ccccc}
1&0&-\frac{y}{2}&0&\frac{i}{2}\theta^+\\
0&1&\frac{x}{2}&0&-\frac{1}{2}\theta^+\\
0&0&1&0&0\\
-\frac{i}{2}\theta^+&\frac{1}{2}\theta^+&\frac{i}{2}\theta^-&1&0\\
0&0&\frac{i}{2}\theta^+&0&1\end{array}
\right)
\eea
and whose inverse is given as
\bea
(E^{-1})_A{}^M=\left(
\begin{array}{ccccc}
1&0&\frac{y}{2}&0&-\frac{i}{2}\theta^+\\
0&1&-\frac{x}{2}&0&\frac{1}{2}\theta^+\\
0&0&1&0&0\\
\frac{i}{2}\theta^+&-\frac{1}{2}\theta^+&
-\frac{i}{2}\theta^-+\frac{1}{4}(x+iy)\theta^+&1&0\\
0&0&-\frac{i}{2}\theta^+&0&1\end{array}
\right)~~~.
\eea
Therefore the supercovariant derivatives are given as
\bea
D_A=(E^{-1})_A{}^M(\frac{1}{i})\partial_M
\eea
whose components are \bref{covosp}.
\par

\section{Global charges }
In this appendix, we give a derivation of (\ref{glocha})
Under the global transformations a group element $g$ is transformed into
$g\to Gg$ with infinitesimal parameters $\varepsilon$
\bea
g^{-1}\delta_\varepsilon g=g^{-1}(G-1)g\equiv \Delta_\varepsilon E^AT_A
=\delta_\varepsilon z^ME_M{}^A~T_A~~.
\eea  
Expression of $\Delta_\varepsilon E^A$'s are obtained as
\bea
g^{-1}(e^{i\varepsilon^+ Q_+}-1)g&=&
i\varepsilon^+[
-i\theta^+ L_++i\left(\theta^--\frac{i}{2}(x+iy)\theta^+\right)L_z
+Q_+-\frac{i}{2}(x+iy)Q_-
]\nn\\
g^{-1}(e^{i\varepsilon^- Q_-}-1)g&=&
i\varepsilon^-[
i\theta^+L_z
+Q_-
]\nn\\
g^{-1}(e^{i\varepsilon^x L_x}-1)g&=&
i\varepsilon^x[L_x-yL_z+\frac{i}{2}\theta^+Q_-
]\nn\\
g^{-1}(e^{i\varepsilon^y L_y}-1)g&=&
i\varepsilon^y[L_y+xL_z-\frac{1}{2}\theta^+Q_-
]\nn\\
g^{-1}(e^{i\varepsilon^\phi L_z}-1)g&=&
i\varepsilon^\phi~L_z\label{Dtrans}~~~.
\eea

The global charges are given by
\bea
\varepsilon\hat{\cal Q}=\delta_\varepsilon z^M(\frac{1}{i})\partial_M=
\Delta_\varepsilon E^A(E^{-1})_A{}^M(\frac{1}{i})\partial_M
~~~.\label{glocharge}
\eea




\vspace{0.5cm}
\begin{thebibliography}{99}
\bibliographystyle{unsrt}
%
\setlength{\itemsep}{0.0in}

\bibitem{OV} 
H.~Ooguri and C.~Vafa,
arXiv:hep-th/0302109, \ 
hep-th/0303063.
\bibitem{Peter}
J.~de Boer, P.~A.~Grassi and P.~van Nieuwenhuizen,
arXiv:hep-th/0302078.
\bibitem{Seiberg} 
N.~Seiberg,
JHEP {\bf 0306}, 010 (2003)
[arXiv:hep-th/0305248].
\bibitem{shomerus} V.~Schomerus,
JHEP {\bf 9906}, 030 (1999)
[arXiv:hep-th/9903205].
\bibitem{SW} 
N.~Seiberg and E.~Witten,
JHEP {\bf 9909}, 032 (1999)
[arXiv:hep-th/9908142].
\bibitem{schwarz}
J.~H.~Schwarz and P.~Van Nieuwenhuizen,
Lett.\ Nuovo Cim.\  {\bf 34} (1982) 21.
\bibitem{Kosinski} 
P.~Kosinski, J.~Lukierski and P.~Maslanka,
arXiv:hep-th/0011053.
\bibitem{Ferrara} 
S.~Ferrara and M.~A.~Lledo,
JHEP {\bf 0005}, 008 (2000)
[arXiv:hep-th/0002084].
\bibitem{Klemm}
 D.~Klemm, S.~Penati and L.~Tamassia,
Class.\ Quant.\ Grav.\  {\bf 20}, 2905 (2003)
[arXiv:hep-th/0104190].
[arXiv:math-ph/9804013].
\bibitem{AIIKKT} 
H.~Aoki, N.~Ishibashi, S.~Iso, H.~Kawai, Y.~Kitazawa and T.~Tada,
Nucl.\ Phys.\ B {\bf 565}, 176 (2000)
[arXiv:hep-th/9908141].
\bibitem{IIKK} N.~Ishibashi, S.~Iso, H.~Kawai and Y.~Kitazawa,
Nucl.\ Phys.\ B {\bf 573}, 573 (2000)
[arXiv:hep-th/9910004].
\bibitem{Seiberg2} N.~Seiberg,
JHEP {\bf 0009}, 003 (2000)
[arXiv:hep-th/0008013].
\bibitem{Grosse} 
H.~Grosse, C.~Klimcik and P.~Presnajder,
Commun.\ Math.\ Phys.\  {\bf 185}, 155 (1997)
[arXiv:hep-th/9507074].
\bibitem{Grosse2}
 H.~Grosse and G.~Reiter,
Jour.\ Geom.\ Phys.\ {\bf 28}, 349 (1998)
\bibitem{HIU22} M. Hatsuda, S. Iso and H. Umetsu, to appear 
\bibitem{Kawai} H.~Kawai, T.~Kuroki and T.~Morita,
arXiv:hep-th/0303210.
\bibitem{sohnius} M.F. Sohnius, Physics Report 128 (1985) 39 - 204
\bibitem{1001} S.J.Gates, M.T.Grisaru, M.Rocek and W. Siegel, Superspace,
Benjamin/Cunnings Publishing Company, Inc. 1983
\bibitem{IKKT} 
N.~Ishibashi, H.~Kawai, Y.~Kitazawa and A.~Tsuchiya,
Nucl.\ Phys.\ B {\bf 498}, 467 (1997)
[arXiv:hep-th/9612115].
\bibitem{supermatrix}
L.~Smolin,
arXiv:hep-th/0006137.
T.~Azuma, S.~Iso, H.~Kawai and Y.~Ohwashi,
Nucl.\ Phys.\ B {\bf 610}, 251 (2001)
[arXiv:hep-th/0102168].
\bibitem{representation} 
A.~Pais and V.~Rittenberg,
J.\ Math.\ Phys.\  {\bf 16}, 2062 (1975)
[Erratum-ibid.\  {\bf 17}, 598 (1976)].
M.~Scheunert, W.~Nahm and V.~Rittenberg,
J.\ Math.\ Phys.\  {\bf 18}, 155 (1977).
M.~Marcu,
J.\ Math.\ Phys.\  {\bf 21}, 1277 (1980).
\bibitem{alek} A.~Y.~Alekseev, A.~Recknagel and V.~Schomerus,
JHEP {\bf 9909}, 023 (1999)
[arXiv:hep-th/9908040].
\bibitem{fuzzy2} J.~Madore,
Class.\ Quant.\ Grav.\  {\bf 9}, 69 (1992).
\bibitem{IKTW} S.~Iso, Y.~Kimura, K.~Tanaka and K.~Wakatsuki,
Nucl.\ Phys.\ B {\bf 604}, 121 (2001)
[arXiv:hep-th/0101102].
\bibitem{Coleman}
S.~R.~Coleman,
``The Magnetic Monopole Fifty Years Later,''
HUTP-82/A032
\bibitem{Balachandran}
A.~P.~Balachandran, S.~Kurkcuoglu and E.~Rojas,
JHEP {\bf 0207}, 056 (2002)
[arXiv:hep-th/0204170].
\bibitem{IW}  E.In$\ddot{\rm o}$n$\ddot{\rm u}$ and E.P.Wigner, Proc. Natl. Acad. Sci. U.S.A.
{\bf 39} (1953) 510,\\
E.In$\ddot{\rm o}$n$\ddot{\rm u}$, 
in 
``{\it Group Theoretical Concepts  and Methods in Elementary Particle Physics}'' ed. F.Gursey (Gordon and Breach, New York, 1964).

\bibitem{HKSpp} M.Hatsuda, K.Kamimura and M.Sakaguchi, 
Nucl. Phys. {\bf B632} (2002) 114, hep-th/0202190;
Nucl. Phys. {\bf B637} (1-3) (2002) 168, hep-th/0204002.
Prog. Theor. Phys. {\bf 109} (2003) 853, hep-th/0106114.
\bibitem{gIW}
M.Hatsuda and M.Sakaguchi,  Phys. Rev. {\bf D66} (2002) 045020, hep-th/0205092;  
Prog. Theor. Phys. {\bf 109} (2003) 853, hep-th/0106114.
\bibitem{az} J. A. de Azcarraga, J.M. Izquiero, M. Picon abd O. Varela, hep-th/0212347.
\bibitem{Klimcik}
C.~Klimcik,
Commun.\ Math.\ Phys.\  {\bf 206}, 587 (1999)
[arXiv:hep-th/9903202].
\bibitem{IKS} 
S.~Iso, D.~Karabali and B.~Sakita,
Phys.\ Lett.\ B {\bf 296}, 143 (1992)
[arXiv:hep-th/9209003].
Nucl.\ Phys.\ B {\bf 388}, 700 (1992)
[arXiv:hep-th/9202012].
A.~Cappelli, C.~A.~Trugenberger and G.~R.~Zemba,
Phys.\ Lett.\ B {\bf 306}, 100 (1993)
[arXiv:hep-th/9303030].
\bibitem{Rey} R.~Britto, B.~Feng and S.~J.~Rey,
arXiv:hep-th/0306215.
\bibitem{BS} N. Berkovits and N. Seiberg, hep-th/0306226

\end{thebibliography}
\end{document}